# Interlinked Cycles for Index Coding:
# Generalizing Cycles and Cliques

Chandra Thapa, Lawrence Ong, and Sarah J. Johnson

## Abstract

We consider a graphical approach to index coding. While cycles have been shown to provide coding gain, cycles and cliques (a specific type of overlapping cycles) have been exploited in existing literature. In this paper, we define a more general form of overlapping cycles, called the interlinked-cycle (IC) structure, that generalizes cycles and cliques. We propose a scheme, called the interlinked-cycle-cover (ICC) scheme, that leverages IC structures in digraphs to construct scalar linear index codes. We characterize a class of infinitely many digraphs where our proposed scheme is optimal over all linear and non-linear index codes. Consequently, for this class of digraphs, we indirectly prove that scalar linear index codes are optimal. Furthermore, we show that the ICC scheme can outperform all existing graph-based schemes (including partial-clique-cover and fractional-local-chromatic number schemes), and a random-coding scheme (namely, composite coding) for certain graphs.

## I. INTRODUCTION

In index coding (introduced by Birk and Kol [1] in 1998), a sender broadcasts messages through a noiseless shared channel to multiple receivers, each knowing some messages a priori, which are known as side information. Side information occurs frequently in many communication networks, e.g., in a web browsers' cache. Knowing the side information of the receivers, the sender can send coded symbols, known as an index code, in such a way that all of the receivers can decode their requested messages using their side information and the received coded symbols. The aim is to find the shortest (optimal) index code. How to optimally design an index code for an arbitrary index-coding instance, a specific index-coding problem characterized by the message sets each receiver wants and knows, is an open problem to date.

In the literature, various approaches have been adopted to solve the index-coding problem. We broadly classify these approaches into four categories: (i) numerical, (ii) Shannon's random coding, (iii) interference alignment, and (iv) graph-based. Numerical approaches include rank minimization over finite fields [2], [3] (which is NP-hard to compute in general [4], [5]), and mathematical optimization programming (semi-definite programming [6], linear programming [7], and integer-linear programming [8]). These approaches do not provide much intuition on the interaction between the side-information configuration and the index codes. Shannon's random coding approaches [9], [10] require infinitely long message packets. Interference-alignment approaches treat index coding as an interference-alignment problem [11], [12], and construct index codes via two alignment techniques, namely one-to-one alignment and subspace alignment. These alignment techniques have no well-defined algorithms to construct index codes for arbitrary index-coding instances. Graph-based approaches [8], [13]–[22] provide intuition on the side-information configurations and index codes. These approaches represent index-coding instances by graphs, and construct index codes as functions of the graphs. These graph-based schemes provide linear (scalar and vector) index codes. Although linear index codes are not always optimal [3], [23], they have simpler encoding and decoding processes.

We classify graph-based approaches into two sub-categories: (i) Maximum distance separable (MDS) code based interference alignment approaches, and (ii) graph structure based approaches. The MDS code based interference alignment approaches construct index codes by treating messages not known to a receiver as interference, and aligning all interference with the help of MDS codes. These approaches include the partial-clique-cover scheme [13] and its fractional version [8], [14], the local-chromatic-number scheme and its fractional version [15], and the partitioned-local-chromatic-number scheme and its fractional version [16]. Graph structure based approaches exploit special graph structures, based on messages known to the receivers that can provide some savings[1] in index-coding instances.

It has been shown that no structure in an acyclic graph can provide any savings [2]. Furthermore, if an arc does not belong to any cycle, then removing it does not change the optimal index code

---

[1]The number of transmissions saved by transmitting coded symbols (coded messages) rather than transmitting uncoded messages is called *savings* of the index code. For savings, greater is better.

[19]–[21]. These observations point to the importance of cycles in index coding. In the literature, only cycles and cliques, a specific combination of overlapping cycles, have been exploited so far in the graph structure based approaches. More precisely, cycles in digraphs are exploited by the cycle-cover scheme [17], [18] and its fractional version [17], and cliques in digraphs are exploited by the clique-cover scheme [13] and its fractional version [7]. Overlapping cycles can provide more savings than considering cycles individually. We take a clique as an example. In a clique, every vertex forms a cycle with any other vertex, and we see overlapping of cycles at every vertex. If we consider only cycles in the clique, we get an index code strictly longer than that by considering the clique. However, not all forms of overlapping cycles are useful, in the sense that they provide more savings than considering cycles and cliques. In this work, we consider a graph structure based approach, and propose structures of overlapping cycles that can be exploited in graphs to provide potentially more savings than the cycle-cover scheme, the clique-cover scheme, and other existing schemes. The proposed structures are called interlinked-cycle (IC) structures, and they generalize cycles and cliques. Furthermore, we define a scheme, called the interlinked-cycle cover (ICC) scheme, that constructs index codes based on IC structures. In addition, we extend the ICC scheme to allow time-sharing among overlapping IC structures, we call the scheme the fractional-ICC scheme. Also, we extend the IC structures and present a scheme that exploits the extended-IC structures, we call the scheme the extended-ICC scheme. Referring to the work by Ong [24], for all index-coding instances up to five receivers (9846 problems) where each receiver requests exactly one message, and each message is requested by exactly one receiver, we find that the ICC scheme (including fractional and extended versions) provides the minimum index codelength for all message alphabet sizes except problems modeled by eight non-isomorphic digraphs (see Remark 9 for the details).

## A. Our contributions

1) We propose a new index-coding scheme (called the ICC scheme) that generalizes the clique-cover scheme and the cycle-cover scheme. The new scheme constructs scalar linear index codes.

2) We characterize a class of digraphs (with infinitely many members) for which the ICC scheme is optimal (over all linear and non-linear index codes). This means scalar linear index codes are optimal for this class of digraphs.

3) For a class of digraphs, we prove that the ICC scheme performs at least as well as the

partial-clique-cover scheme. We conjecture that the result is valid in general. Furthermore, we present a class of digraphs where the additive gap between these two schemes grows linearly with the number of vertices in the digraph.

4) For a class of digraphs, we prove that the ICC scheme performs at least as well as the fractional-local-chromatic-number scheme. Moreover, we present a class of digraphs where the additive gap between these two schemes grows linearly with the number of vertices in the digraph.

5) We show that the ICC scheme can outperform all of the existing graph-based schemes and the composite-coding scheme in some examples.

6) We extend the ICC scheme to the fractional-ICC scheme. This modified scheme time-shares multiple IC structures, and constructs vector linear index codes that can be, for certain digraphs, shorter than the scalar linear index codes obtained from the ICC scheme.

## II. DEFINITIONS AND BACKGROUND

### A. Problem formulation

Consider a transmitter that wants to transmit $N$ messages $X = \{x_1, x_2, \ldots, x_N\}$ to $N$ receivers $\{1, 2, \ldots, N\}$ in a *unicast* message setting, meaning that each message is requested by only one receiver, and each receiver requests only one message. Without loss of generality, let each receiver $i$ request message $x_i$, and possess side information $S_i \subseteq X \setminus \{x_i\}$. This problem can be described by a digraph $D = (V(D), A(D))$, where $V(D) = \{1, 2, \ldots, N\}$, the set of vertices in $D$, represents the $N$ receivers. An arc $(i \rightarrow j) \in A(D)$ exists from vertex $i$ to vertex $j$ if and only if receiver $i$ has packet $x_j$ (requested by receiver $j$) in its side information. The set of the side information of a vertex $i$ is $S_i \triangleq \{x_j : j \in N_D^+(i)\}$, where $N_D^+(i)$ is the out-neighborhood of $i$ in $D$. Let $X$ and all $S_i$ be ordered sets, where the ordering can be arbitrary but fixed.

*Definition 1 (Index code):* Suppose $x_i \in \{0, 1\}^t$ for all $i$ and some integer $t \geq 1$, i.e., each message consists of $t$ bits. Given an index-coding problem modeled by $D$, an index code $(\mathscr{F}, \{\mathscr{G}_i\})$ is defined as follows:

1) An encoding function for the source, $\mathscr{F} : \{0, 1\}^{N \times t} \rightarrow \{0, 1\}^p$, which maps $X$ to a $p$-bit index code for some positive integer $p$.

2) A decoding function $\mathscr{G}_i$ for every receiver $i$, $\mathscr{G}_i : \{0, 1\}^p \times \{0, 1\}^{|S_i| \times t} \rightarrow \{0, 1\}^t$, that maps the received index code $\mathscr{F}(X)$ and its side information $S_i$ to the requested message $x_i$.

*Definition 2 (Broadcast rate or index codelength):* The broadcast rate of an index code $(\mathscr{F}, \{\mathscr{G}_i\})$ is the number of transmitted bits per received message bits at every user, or equivalently the number of transmitted coded symbols (each of $t$ bits). This is denoted by $\ell(D) \triangleq \frac{p}{t}$, and is also referred to as the index codelength.

*Definition 3 (Optimal broadcast rate):* The optimal broadcast rate for a given index coding problem $D$ with $t$-bit messages is $\beta_t(D) \triangleq \min_{\mathscr{F}} \ell(D)$, and the optimal broadcast rate over all $t$ is defined as $\beta(D) \triangleq \inf_t \beta_t(D)$.

*Remark 1:* For broadcast rates, the minimum is optimal.

*Definition 4 (Maximum acyclic induced sub-digraph):* For a digraph $D$, an induced acyclic sub-digraph formed by removing the minimum number of vertices in $D$, is called a maximum acyclic induced sub-digraph (MAIS). The order of the MAIS is denoted as $\mathsf{MAIS}(D)$.

It has been shown that for any digraph $D$ and any message length of $t$-bits, $\mathsf{MAIS}(D)$ lower bounds the optimal broadcast rate [2],

$$\mathsf{MAIS}(D) \leq \beta(D) \leq \beta_t(D). \tag{1}$$

## B. Some existing schemes

In this sub-section, we describe the clique-cover, the cycle-cover and the partial-clique-cover schemes in detail. These schemes provide some basic intuitions about our proposed ICC scheme. Throughout this paper, "disjoint" refers to "vertex-disjoint" unless stated otherwise.

*Definition 5 (Clique):* A clique is a sub-digraph where each vertex has an out-going arc to every other vertex in that sub-digraph.

*Definition 6 (Clique-covering number, $\ell_{\mathsf{CL}}(D)$):* The clique-covering number is the minimum number of cliques partitioning a digraph $D$ (over all partitions) such that if each clique is denoted as $\rho_i$ for $i \in \{1, \ldots, \ell_{\mathsf{CL}}(D)\}$, then $\bigcup_{\forall i} V(\rho_i) = V(D)$, $V(\rho_i) \cap V(\rho_j) = \emptyset$, $\forall i, j \in \{1, \ldots, \ell_{\mathsf{CL}}(D)\}$ and $i \neq j$ (here a vertex is a clique of size one).

The clique-covering number is equal to the chromatic number of the underlying undirected graph of the complement digraph, i.e., $\ell_{\mathsf{CL}}(D) = \chi(U_{\bar{D}})$, where $U_{\bar{D}}$ is the underlying undirected graph of $\bar{D}$, $\bar{D}$ is the complement digraph of $D$, and $\chi(D)$ denotes the chromatic number of $D$.

*Definition 7 (Clique-cover scheme):* The clique-cover scheme finds a set of disjoint cliques that provides the clique-covering number, and constructs an index code in which the coded symbol for each of the disjoint cliques is the bit-wise XOR of messages requested by all of the vertices in that clique.

*Remark 2:* Any clique with $n$ vertices permits a savings of $n - 1$.

The clique-cover scheme achieves the following rate:

*Proposition 1 (Birk and Kol [13]):* The optimal broadcast rate of an index coding instance is upper bounded by the clique cover number, i.e., $\beta(D) \leq \beta_t(D) \leq \ell_{\mathsf{CL}}(D), \ \forall t$.

*Definition 8 (Path and cycle):* A *path* consists a sequence of distinct (except possibly the first and last) vertices, say $1, 2, \ldots, M$, and an arc $(i \rightarrow i + 1)$ for each consecutive pair of vertices $(i, i + 1)$ for all $i \in \{1, 2, \ldots, M - 1\}$. We represent this path from the vertex 1 to the vertex $M$ as $\langle 1, 2, \ldots, M \rangle$. Here 1 is the *first vertex* and $M$ is the *last vertex*. A path having (i) at least two distinct vertices, and (ii) the same first and last vertex is a *cycle*. Moreover, in any cycle, the in-degree and the out-degree of any vertex are both exactly one.

*Definition 9 (Cycle-covering number, $\ell_{\mathsf{CY}}(D)$):* The difference between the total number of vertices in $D$ and the maximum number of disjoint cycles in $D$ is the cycle-covering number.

*Definition 10 (Cycle-cover scheme):* The cycle-cover scheme finds a set of disjoint cycles in $D$ that provides the cycle-covering number, and constructs an index code that has (i) coded symbols for each disjoint cycle (for a cycle $\langle 1, 2, \ldots, M, 1 \rangle$, a set of coded symbols are $\{x_1 \oplus x_2, \ x_2 \oplus x_3, \ldots, \ x_{M-1} \oplus x_M\}$), and (ii) uncoded messages which are requested by those vertices not included in any of the disjoint cycles in $D$.

*Remark 3:* Any cycle permits a savings of one.

The cycle-cover scheme achieves the following rate:

*Proposition 2 ( [13], [17], [18]):* The optimal broadcast rate of an index-coding problem is upper bounded by the cycle-covering number, i.e., $\beta(D) \leq \beta_t(D) \leq \ell_{\mathsf{CY}}(D), \ \forall t$.

*Definition 11 (Partial clique):* A sub-digraph $D_i = (V(D_i), A(D_i))$ is a $\kappa(D_i)$-partial clique, where $\kappa(D_i) = |V(D_i)| - \delta^+(D_i) - 1$ and $\delta^+(D_i)$ is the minimum out-degree of $D_i$.

*Definition 12 (Partial-clique number, $\ell_{\mathsf{PC}}(D)$):* If $D_1, D_2, \ldots, D_m$, for some positive integer $m$, partition a digraph $D$ such that $\bigcup_{\forall i \in \{1,2,\ldots,m\}} V(D_i) = V(D)$ and $V(D_i) \cap V(D_j) = \emptyset$, $\forall i, j \in \{1, 2, \ldots, m\}$ and $i \neq j$, then

$$\ell_{\mathsf{PC}}[D_1, D_2, \ldots, D_m] = \sum_{i=1}^{m} (\kappa(D_i) + 1),$$

and the partial-clique number of the digraph is

$$\ell_{\mathsf{PC}}(D) = \min_{D_1, D_2, \ldots, D_m} \ell_{\mathsf{PC}}[D_1, D_2 \ldots, D_m], \tag{2}$$

where the minimum is taken over all partitions.

*Definition 13 (Partial-clique-cover scheme):* The partial-clique-cover scheme finds a set of disjoint partial cliques in $D$ that provides the partial-clique number, and constructs an index code that has (i) $\kappa(D_i) + 1$ coded symbols for each disjoint $\kappa(D_i)$-partial clique with $|V(D_i)| > 1$ (a partial clique uses MDS codes to generate coded symbols), and (ii) an uncoded message for each disjoint $\kappa(D_i)$-partial clique with $|V(D_i)| = 1$.

The partial-clique-cover scheme achieves the following rate:

*Proposition 3 (Birk and Kol [13]):* The optimal broadcast rate of an index coding instance is upper bounded by the partial-clique number, i.e.,

$$\beta(D) \leq \ell_{\mathsf{PC}}(D). \tag{3}$$

*Remark 4:* The partial-clique-cover scheme performs at least as well as the cycle-cover and the clique-cover schemes, i.e.,

$$\ell_{\mathsf{PC}}(D) \leq \min \left\{ \ell_{\mathsf{CL}}(D), \ell_{\mathsf{CY}}(D) \right\}. \tag{4}$$

This is because the partial-clique-cover scheme includes the cycle-cover scheme or the clique-cover scheme as a special case. By definition, a clique is a $0$-partial clique, and a cycle with $n$ vertices is a $(n-2)$-partial clique. Despite the fact that the partial-clique-cover scheme uses MDS codes, which require sufficiently large message length $t$ in general, to construct index codes, one can find MDS codes for any cycle and any clique for any $t$.

*Remark 5:* The clique-cover, the cycle-cover and the partial-clique-cover schemes provide scalar linear index codes. We can also construct vector linear index codes by time-sharing all possible cliques, cycles, partial cliques in their respective schemes, and these are called the fractional versions of those schemes. The fractional version can strictly decrease the broadcast rates (over the non-fractional version) for some digraphs, e.g., a 5-cycle [7].

## III. MOTIVATING OVERLAPPING CYCLES

We present an example that illustrates the importance of overlapping cycles in index coding. Consider the digraph $D_1$ in Fig. 1a. In $D_1$, the cycles $\langle 1, 2, 1 \rangle$ and $\langle 1, 4, 3, 5, 1 \rangle$ overlap at vertex 1, and some cycles similarly overlap at vertices 2 and 3. Note that $\mathsf{MAIS}(D_1) = 3 \leq \beta(D_1)$. Index codelengths for $D_1$ by existing graph-based schemes (some schemes require a sufficiently large $t$) are depicted in Table I.

Among all existing schemes listed in Table I, only the composite-coding scheme (which requires infinitely long message length) can achieve $\beta(D_1)$. However, there exists a scalar linear

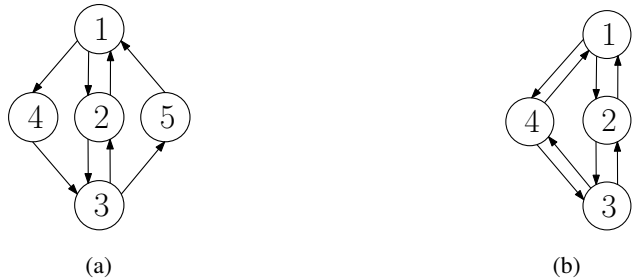

(a)                                                (b)

Fig. 1. Two digraphs with overlapping cycles, (a) $D_1$, and (b) $D_2$.

Index codelengths for the digraph $D_1$ in Fig. 1a from existing schemes

| Schemes | Index codelength |
|---|:---:|
| Clique cover [13] | 4 |
| Fractional-clique cover [7] | 4 |
| Cycle cover [17], [18] | 4 |
| Fractional-cycle cover [17] | 3.5 |
| Partial-clique cover [13] | 4 |
| Fractional-partial-clique cover [14] | 3.5 |
| Local-chromatic number [15] | 4 |
| Local time sharing [14], $b_{\mathsf{LTS}}$ | 3.5 |
| Asymmetric coding [14], $b(\mathscr{R}_{\mathsf{LTS}})$ | 3.5 |
| Composite coding [9] | 3 |

index code of length three, i.e., $\{x_1 \oplus x_2 \oplus x_3, \ x_4 \oplus x_3, \ x_5 \oplus x_1\}$, which is not found by these existing index-coding schemes. Later in this paper, we will show how to obtain this code by considering overlapping cycles.

In some digraphs with overlapping cycles, the optimal broadcast rate can be obtained by the cycle-cover scheme (which only codes on disjoint cycles). The digraph $D_2$ in Fig. 1b gives one of these cases, where the optimal broadcast rate $\beta(D_2) = 2$ is achieved by the cycle-cover scheme, $\ell_{\mathsf{CY}}(D_2) = 2$. For these digraphs, there is no benefit considering overlapping cycles.

This paper explores structures with overlapping cycles having index codes with length strictly shorter than that obtained from the cycle-cover scheme (possibly other existing schemes as well). We will identify a class of such structures, and call them IC structures. We will also show that exploiting IC structures can strictly outperform the composite-coding scheme for some digraphs.

## IV. Understanding IC structures and construction of index codes

In this section, we provide an informal description of an IC structure and code construction. This provides an insight into IC structures.

The main idea behind our definition of an IC structure is the existence of a set of vertices with the following property: For any pair of vertices in the set, there exists a cycle containing these two vertices, and no other vertices in the set. We call such a vertex set an *inner vertex set*, and its vertices are called *inner vertices* (see Section V-A for the formal definition). The remaining vertices that are not inner vertices are called *non-inner vertices*. Now we illustrate IC structures with an example. For the digraph in Fig. 1a, $\{1, 2, 3\}$ is an inner vertex set. This is because the cycle $\langle 1, 2, 1 \rangle$ includes the vertex pair $(1, 2)$, the cycle $\langle 2, 3, 2 \rangle$ includes the vertex pair $(2, 3)$, and the cycle $\langle 1, 4, 3, 5, 1 \rangle$ includes the vertex pair $(1, 3)$. Consequently, the vertices 1, 2 and 3 are interlinked by cycles that are not disjoint (e.g., the vertex 1 is in cycles $\langle 1, 2, 1 \rangle$ and $\langle 1, 4, 3, 5, 1 \rangle$), and such cycles are called interlinked cycles. In Section V-A, we will formally define an IC structure to be a sub-digraph consisting of paths among the inner vertices rather than cycles between different inner vertex pairs. However, the two definitions are linked by the observation that two disjoint (except for the end vertices) paths $\langle A, \ldots, B \rangle$ and $\langle B, \ldots, A \rangle$ form a cycle.

The construction of an index code for the IC structure is as follows: (i) One codeword symbol is formed by the XOR of messages requested by all of the inner vertices (e.g., for the digraph in Fig. 1a, the vertices 1, 2 and 3 are inner vertices, so one of the coded symbols is $x_1 \oplus x_2 \oplus x_3$), and (ii) for each of the remaining vertices (non-inner vertices), a codeword symbol is formed by the XOR of the message requested by the vertex and the messages requested by all of its out-neighborhood vertices (e.g., for the digraph in Fig. 1a, vertices 4 and 5 are non-inner vertices, so the corresponding coded symbols are $x_4 \oplus x_3$ and $x_5 \oplus x_1$). Thus the index code for the digraph in Fig. 1a (which is also an IC structure) is $\{x_1 \oplus x_2 \oplus x_3, \ x_4 \oplus x_3, \ x_5 \oplus x_1\}$. Now the decoding process is as follows: Vertex 2 decodes its requested message $x_2$ from the coded symbol $x_1 \oplus x_2 \oplus x_3$ because it has $x_1$ and $x_3$ in its side information. Vertices 4 and 5 decode their requested messages $x_4$ and $x_5$ from the coded symbols $x_4 \oplus x_3$ and $x_5 \oplus x_1$ respectively because vertex 4 has $x_3$ in its side information, and vertex 5 has $x_1$ in its side information. Vertex 1 decodes its requested message $x_1$ from $(x_1 \oplus x_2 \oplus x_3) \oplus (x_4 \oplus x_3) = x_1 \oplus x_2 \oplus x_4$ because it has $x_2$ and $x_4$ in its side information. Vertex 3 decode its requested message $x_3$ from $(x_1 \oplus x_2 \oplus x_3) \oplus (x_5 \oplus x_1) = x_2 \oplus x_5 \oplus x_3$ because it has $x_2$ and $x_5$ in its side information.

## V. An IC structure and construction of the Index Code

In this section, we provide the formal definition of an IC structure and code construction.

### A. Definition of an IC structure

In a digraph $D$, consider a structure having the following property:

- It has an *inner vertex* set denoted as $V_I$. Without loss of generality, we consider $V_I = \{1, 2, \ldots, K\}$. The vertices in $V_I$ are called inner vertices. For any ordered pair of inner vertices $(i, j) \in V_I$ and $i \neq j$, there is a path from $i$ to $j$, and the path does not include any other vertex in $V_I$ except the first and the last vertices.

Due to the existence of paths between any vertex pair in $V_I$, for each inner vertex $i \in V_I$, we can always find a *directed rooted tree* in $D$, denoted by $T_i$, where inner vertex $i$ is the root vertex, and all other inner vertices $V_I \setminus \{i\}$ are the leaves (see Fig. 2a). The trees may be non-unique. If the union of $K$ selected trees, each one rooted at a unique inner vertex (i.e., $\bigcup_{\forall i \in V_I} T_i$ (see Fig. 2b)), satisfies two conditions (to be defined shortly), we call it an interlinked-cycle structure (denoted as a $K$-IC sub-digraph: $D_K = (V(D_K), A(D_K))$, where $V(D_K) \subseteq V(D)$, $A(D_K) \subseteq A(D)$, and $V_I = \{1, 2, \ldots, K\}$). Now we define a type of cycle and a type of path.

*Definition 14 (I-path, I-cycle):* A path in which only the first and the last vertices are from $V_I$, and they are distinct, is an *I-path*. If the first and the last vertices are the same, then it is an *I-cycle*. In other words, an I-cycle is a directed cycle containing exactly one inner vertex.

The conditions for an IC structure are as follows:

1) *Condition 1:* There is no I-cycle in $D_K$.

2) *Condition 2:* For all ordered pairs of inner vertices $(i, j)$, $i \neq j$, there is only one I-path from $i$ to $j$ in $D_K$ (this condition implies that in $D_K$, for any two inner vertices, there is exactly one cycle containing them and no other inner vertices).

These two conditions are necessary for our code construction described in the following section (for more discussions, see Examples 2 and 3).

### B. Construction of an index code for a $K$-IC structure

We propose the following coded symbols for a $K$-IC structure $D_K$ with $|V(D_K)| = N'$:

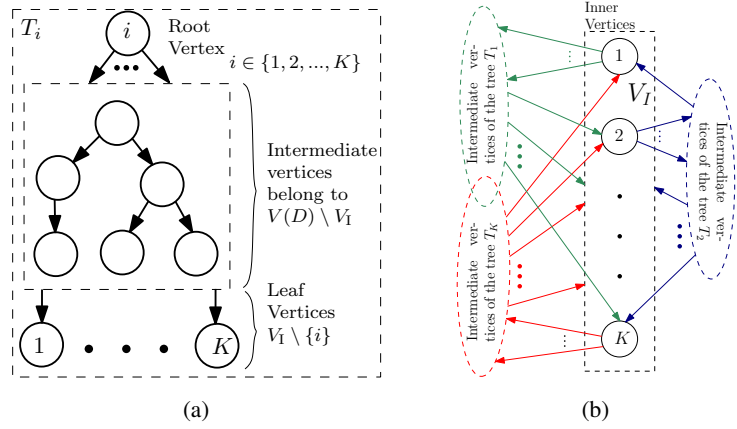

Fig. 2. (a) Outline of a tree $T_i$ which has the root vertex $i \in V_I$, all vertices in $V_I \setminus \{i\}$ as the leaf vertices, and some or all vertices in $V(D) \setminus V_I$ (i.e., the non-inner vertex set) as intermediate vertices (between the root and the leaves), and b) outline of an IC structure, where different trees $T_i$ (shown in different colors) can share intermediate vertices.

1) A coded symbol obtained by the bitwise XOR (denoted by $\oplus$) of messages (each of $t$ bits) requested by all vertices of the inner vertex set $V_I$, i.e.,

$$w_I \triangleq \bigoplus_{i=1}^{K} x_i. \tag{5}$$

2) For each non-inner vertex $j \in \{K+1, K+2, \ldots, N'\}$, a coded symbol obtained by the bitwise XOR of the message requested by $j$ with the messages requested by its out-neighborhood vertices in $D_K$, i.e.,

$$w_j \triangleq x_j \oplus \bigoplus_{q \in N_{D_K}^+(j)} x_q. \tag{6}$$

Denote this code constructed for the $K$-IC structure by $W \triangleq \{w_I, w_j : K+1 \leq j \leq N'\}$. The total number of coded symbols in $W$ is

$$\ell(D_K) = N' - K + 1. \tag{7}$$

*Remark 6:* For a given $D_K$, encoding $W$ requires at most $t \times \left( K - 1 + \sum_{i \in V(D_K) \setminus V_I} |N_{D_K}^+(i)| \right)$ bit-wise XOR operations where $t$ is the number of bits in each message symbol.

*Proposition 4:* The code $W$ constructed for a $K$-IC structure is an index code.

*Proof:* Refer to Appendix A. ∎

The following example illustrates the definition of an IC structure.

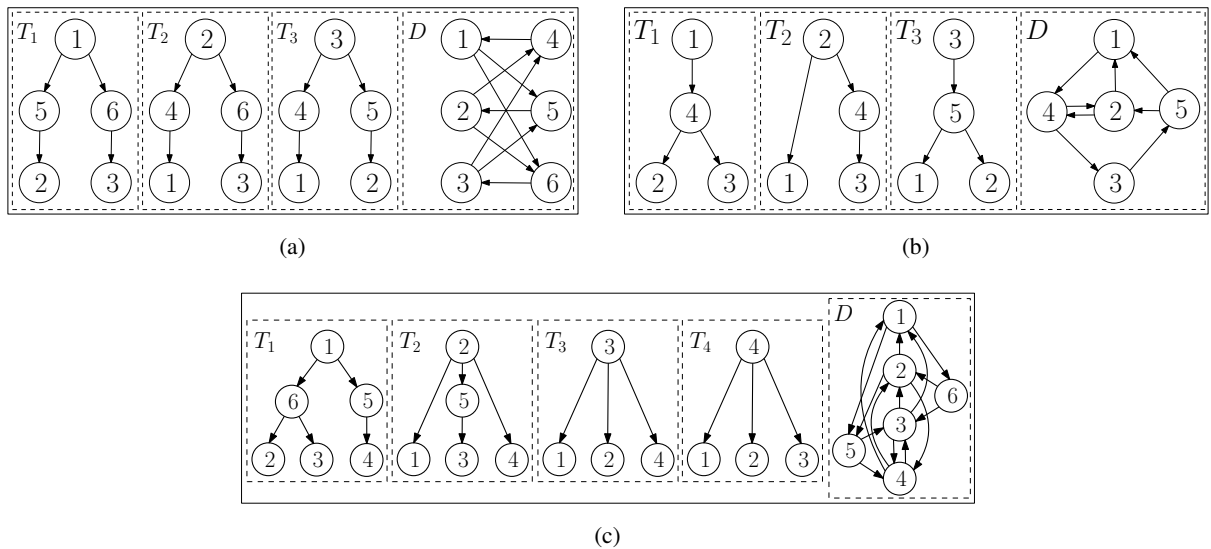

(c)

Fig. 3. (a) A digraph $D$ that is also an IC structure along with its trees $T_1$, $T_2$ and $T_3$, (b) a digraph $D$ along with its trees $T_1$, $T_2$ and $T_3$, has an I-cycle at vertex 2, and (c) a digraph $D$ along with its trees $T_1$, $T_2$, $T_3$ and $T_4$, has two I-paths at vertex 3 from vertex 1.

*Example 1:* Consider the digraph $D$ in Fig. 3a where we have identified the inner-vertex set $V_I = \{1, 2, 3\}$. For vertices 1, 2 and 3, we can always find the directed rooted trees $T_1$, $T_2$ and $T_3$. The union of these three trees form a digraph, denoted $D_K$, with the following: (i) There is no I-cycle in $D_K$, and (ii) the vertex 1 has only one I-path to each of the vertices 2 and 3 (i.e., $\langle 1, 5, 2 \rangle$ and $\langle 1, 6, 3 \rangle$), and so do vertices 2 and 3. Thus $D_K$ is a 3-IC structure with $K = 3$.

The following examples demonstrate why we impose two conditions, viz., *Condition* 1 and *Condition* 2, for the definition of an IC structure.

*Example 2 (Why no I-cycle in $D_K$?):* Consider the digraph $D$ in Fig. 3b where we have identified $V_I = \{1, 2, 3\}$. The digraph has an I-cycle at vertex 2, i.e., $\langle 2, 4, 2 \rangle$. Suppose that we consider $D$ to be an IC structure. We construct a code for $D$ from our proposed code construction; $w_I = x_1 \oplus x_2 \oplus x_3$, $w_4 = x_4 \oplus x_2 \oplus x_3$ and $w_5 = x_5 \oplus x_2 \oplus x_1$. We can verify that vertex 2 with side information $\{x_1, x_4\}$ cannot decode its requested message from the code. This shows that our proposed scheme does not provide an index code for this case. In fact, $\mathsf{MAIS}(D) = 4$, thus no index code of length three exists for $D$. Moreover, the cycle-cover scheme is optimal here.

*Example 3 (Why only one I-path for any ordered pair of the vertices $(i, j) \in V_I$, $i \neq j$, is required in $D_K$?):* Consider the digraph $D$ in Fig. 3c where we have identified $V_I = \{1, 2, 3, 4\}$. The digraph has two I-paths from vertex 1 to vertex 3, viz., $\langle 1, 5, 3 \rangle$ and $\langle 1, 6, 3 \rangle$. Suppose that we consider $D$ as an IC structure. We construct a code for $D$ from our proposed code construction; $w_1 = x_1 \oplus x_2 \oplus x_3 \oplus x_4$, $w_5 = x_5 \oplus x_3 \oplus x_4$ and $w_6 = x_6 \oplus x_2 \oplus x_3$. We can verify that vertex 1 with side information $\{x_5, x_6\}$ cannot decode its requested message from the code. This shows that our proposed scheme does not provide an index code for this case.

## VI. Main Results

In this section, firstly, we define the interlinked-cycle cover (ICC) scheme. Secondly, we prove that the proposed ICC scheme includes the cycle-cover scheme and the clique-cover scheme as special cases. Finally, for a class of digraphs (which includes infinitely many digraphs), we prove that our proposed scheme is optimal, i.e., it achieves the optimal broadcast rates of those digraphs.

*Definition 15 (*ICC *scheme):* For any digraph $D$, the ICC scheme finds a set of disjoint IC structures covering $D$. It then codes each of these IC structures $D_K$ using the code construction described in Section V-B. Note that when $|V(D_K)| = 1$, we simply send the requested message uncoded.

Now we present a main result of this paper. It is best expressed in terms of *savings*, defined as follows:

*Definition 16 (Savings):* The number of transmissions saved by transmitting coded symbols (coded messages) rather than transmitting uncoded messages (i.e., $N - \ell(D)$), is called the *savings* of the index code. It is denoted as $S(D)$.

*Theorem 1:* For any digraph D, an index code of length $\ell_{ICC}(D) = N - \sum_{i=1}^{\psi}(K_i - 1)$ can be achieved by the ICC scheme, where $(K_i - 1)$ is the savings in each disjoint $K_i$-IC structure, and $\psi$ is the number of disjoint IC structures in $D$.

*Proof:* Consider a $K$-IC structure $D_K$ with $N'$ number of vertices. It follows from (7) that the total savings achieved by the ICC scheme in a $D_K$ is

$$N' - \ell_{ICC}(D_K) = N' - (N' - K + 1) = K - 1. \tag{8}$$

For any digraph $D$ containing $\psi$ disjoint IC structures, a savings of $K_i - 1$ is obtained in each $D_{K_i}$ (from (8)), where $i \in \{1, \ldots, \psi\}$. Now the summation of savings in all disjoint IC structures provides the total savings, i.e., $\sum_{i=1}^{\psi}(K_i - 1)$. Hence, $\ell_{ICC}(D) = N - \sum_{i=1}^{\psi}(K_i - 1)$. ∎

*Remark 7 (Complexity issue):* IC structures found by the ICC scheme are not unique. So, finding the best $\ell_{ICC}(D)$ involves optimizing over all choices of disjoint IC structures in $D$, and this may require high time complexity. Refer to Appendix B for an algorithm that exhaustively searches IC structures in $D$ and computes $\ell_{\mathsf{ICC}}(D)$. The complexity can be reduced to some extent in the following way: We partition the digraph into sub-digraphs taking critical arcs[2] [19]–[21] into account, and then search for IC structures in each of the sub-digraphs.

## A. The IC C scheme includes the cycle-cover and the clique-cover schemes as special cases

*Theorem 2:* The ICC scheme includes the cycle-cover and the clique-cover schemes as special cases.

*Proof:* The proof is divided into two parts. In the first part, we prove that the clique-cover scheme is a special case of the ICC scheme, and in the second part, we prove that the cycle-cover scheme is a special case of the ICC scheme.

(Part 1) Consider a clique with $N' \geq 1$ vertices, and without loss of generality, assume that the vertices are $1, 2, \ldots, N'$. We choose $V_{\mathrm{I}} = \{1, 2, \ldots, N'\}$ to be an inner vertex set. As there is exactly one I-path between every ordered pair of vertices, and no I-cycles (indeed no non-inner vertices at all). The clique is a $N'$-IC structure. Now for the clique, (i) the clique-cover scheme constructs an index code $\{x_1 \oplus \cdots \oplus x_{N'}\}$ of length one, and (ii) the ICC scheme constructs an index code $\{w_{\mathrm{I}} = x_1 \oplus \cdots \oplus x_{N'}\}$ of length one. The index codes from both of the schemes are the same. In fact, for any digraph, if we consider the ICC scheme exploiting only $N'$-IC structures with $|V_{\mathrm{I}}| = N'$, i.e., disjoint cliques in digraphs, then this is simply the clique-cover scheme.

(Part 2) Consider a cycle with $N' \geq 2$ number of vertices, and without loss of generality, assume that the vertices are $1, 2, \ldots, N'$. Consider $V_{\mathrm{I}} = \{i, j\}$ for any $i, j \in \{1, 2, \ldots, N'\}$ and $i \neq j$. Vertex $i$ has only one I-path to $j$ and vice versa, and there is no I-cycle. Thus a cycle is a 2-IC structure. Now we choose $i = N'-1$ and $j = N'$, so $V_{\mathrm{I}} = \{N'-1, N'\}$. For the cycle, (i) the cycle-cover scheme constructs an index code $\{x_1 \oplus x_2, \ x_2 \oplus x_3, \ldots, \ x_{N'-1} \oplus x_{N'}\}$ of length $N'-1$, and (ii) the ICC scheme constructs an index code $\{w_{\mathrm{I}}, w_j : 1 \leq j \leq N'-2\}$ of length $N'-1$, where $w_{\mathrm{I}} = x_{N'-1} \oplus x_{N'}$ and $\{w_j : 1 \leq j \leq N'-2\} = \{x_1 \oplus x_2, \ x_2 \oplus x_3, \ldots, x_{N'-2} \oplus x_{N'-1}\}$. The index codes from both of the schemes are the same. In fact, for any digraph, if we consider the ICC scheme exploiting only 2-IC structures, i.e., disjoint cycles in digraphs, then this is simply the cycle-cover scheme. ∎

---

[2] An arc in a digraph is said to be critical if the removal of that arc from the digraph strictly increases the optimal broadcast rate.

*Corollary 1:* The optimal broadcast rate of an index-coding instance, with $t$-bit messages for any $t \geq 1$, is upper bounded by the index codelength obtained from the ICC scheme, and this upper bound is at least as tight as the cycle-covering number and the clique-covering number, i.e.,

$$\beta(D) \leq \beta_t(D) \leq \ell_{\mathsf{ICC}}(D) \leq \min\{\ell_{\mathsf{CL}}(D), \ell_{\mathsf{CY}}(D)\}, \ \forall t. \tag{9}$$

*Proof:* From Theorem 2, the cycle-cover and the clique-cover schemes are special cases of the ICC scheme, thus $\ell_{\mathsf{ICC}}(D) \leq \min\{\ell_{\mathsf{CL}}(D), \ell_{\mathsf{CY}}(D)\}$. From Theorem 1, we have $\beta_t(D) \leq \ell_{\mathsf{ICC}}(D)$, and from the definition of $\beta(D)$, we have $\beta(D) \leq \beta_t(D)$. Therefore, $\beta(D) \leq \beta_t(D) \leq \ell_{\mathsf{ICC}}(D) \leq \min\{\ell_{\mathsf{CL}}(D), \ell_{\mathsf{CY}}(D)\}$. These three schemes are valid for one-bit messages. By concatenating the index codes, we get (9) for any number of bits. ∎

*Remark 8:* $\ell_{\mathsf{CL}}(D)$ and $\ell_{\mathsf{CY}}(D)$ are lower bounded by both $\ell_{\mathsf{ICC}}(D)$ (from Corollary 1) and $\ell_{\mathsf{PC}}(D)$ (from (3)). A difference between these two bounds is that the $\ell_{\mathsf{ICC}}(D)$ bound is valid for all $t$, whereas the $\ell_{\mathsf{PC}}(D)$ bound is valid only for sufficiently large $t$.

### B. The ICC scheme is optimal for a class of digraphs

*Theorem 3:* For messages of any $t \geq 1$ bits, if a $K$-IC structure $D_K$ has

- (Case 1) no cycle containing only non-inner vertices (for example, refer to the digraph $D_1$ in Fig. 1a), or
- (Case 2) exactly $M \geq 1$ cycles containing only non-inner vertices and these cycles are all disjoint, and we can partition the inner-vertex set $V_1$ into $M + 1$ subsets such that each of them forms a disjoint IC structure (possibly by including some non-inner vertices) of case 1, and all of the IC structures are also disjoint from the $M$ cycles among non-inner vertices (for example, refer to the digraph in Fig. 4), then the scalar linear index code given by the ICC scheme is optimal for $D_K$, i.e., $\ell_{\mathsf{ICC}}(D_K) = \beta(D_K) = \beta_t(D_K)$.

*Proof:* Refer to Appendix C. ∎

*Proposition 5:* For messages of any $t \geq 1$ bits and all $K$-IC structures with $K \in \{1, 2, 3\}$, the scalar linear index code given by the ICC scheme is optimal, i.e., $\ell_{\mathsf{ICC}}(D_K) = \beta(D_K) = \beta_t(D_K)$.

*Proof:* Refer to Appendix D. ∎

For all $K$-IC structures that we have constructed (including $K$-IC structures not satisfying Theorem 3), the scalar linear index code given by the ICC scheme is optimal. We conjecture that this holds in general.

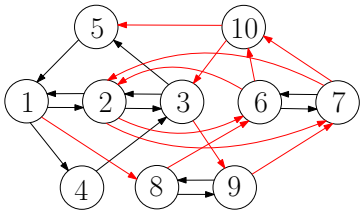

Fig. 4. A 5-IC structure $D_{K=5}$ with $V_I = \{1, 2, 3, 6, 7\}$. For this structure, an index code provided by the ICC scheme is $\{x_1 \oplus x_2 \oplus x_3 \oplus x_6 \oplus x_7, \ x_4 \oplus x_3, \ x_5 \oplus x_1, \ x_8 \oplus x_6 \oplus x_9, \ x_9 \oplus x_7 \oplus x_8, \ x_{10} \oplus x_3 \oplus x_5\}$, which is of length $\ell_{\mathsf{ICC}}(D_{K=5}) = 6$. Furthermore, one can verify that $\mathsf{MAIS}(D_{k=5}) = 6$. In this structure, one can find a cycle $\langle 8, 9, 8 \rangle$ that contains only non-inner vertices (viz., vertices 8 and 9). There is only one cycle of this kind in the structure. So, we get $M = 1$. Now we can partition the inner vertex set $V_I$ into $M+1$, i.e., 2 subsets, viz., $\{1, 2, 3\}$ and $\{6, 7\}$ such that each of the subsets (possibly by including some non-inner vertices) forms a disjoint IC structure (which are also disjoint from the cycle $\langle 8, 9, 8 \rangle$). The first IC structure $D_{K=3}$ is formed by vertices in $\{1, 2, 3, 4, 5\}$ with inner-vertex set $\{1, 2, 3\}$, and the second IC structure $D_{K=2}$ is formed by vertices in $\{6, 7\}$ with inner-vertex set $\{6, 7\}$ (one can find these partitions by removing red arcs from $D_{K=5}$). Furthermore, $D_{K=3}$ and $D_{K=2}$ are two IC structures with no cycle containing only non-inner vertices (i.e., these are IC structures of case 1 of the Theorem 3). Note that any single vertex is also an IC structure with $K = 1$. Now by encoding over the partitions in $D_{K=5}$ (structures remaining after removing all red arcs), the ICC scheme constructs an index code $\{x_1 \oplus x_2 \oplus x_3, \ x_5 \oplus x_1, \ x_4 \oplus x_3, \ x_6 \oplus x_7, \ x_8 \oplus x_9, \ x_{10}\}$ of the length six that is equal to $\ell_{\mathsf{ICC}}(D_{K=5})$.

*Conjecture 1:* For messages of any $t \geq 1$ bits and any $K$-IC structure $D_K$, the scalar linear index code given by the ICC scheme is optimal, i.e., $\ell_{\mathsf{ICC}}(D_K) = \beta(D_K) = \beta_t(D_K)$.

## VII. Comparison with existing schemes

In this section, for a class of digraphs (to be defined shortly), we compare the ICC scheme to the partial-clique-cover scheme, and prove that $\ell_{\mathsf{ICC}}(D) \leq \ell_{\mathsf{PC}}(D)$. We conjecture that $\ell_{\mathsf{ICC}}(D) \leq \ell_{\mathsf{PC}}(D)$ is valid in general. Furthermore, we present a family of digraphs where the additive gap (on the index codelength) between these two schemes grows linearly with the total number of the vertices in the digraph. Secondly, we show all of the above for the fractional-local-chromatic-number scheme instead of the partial-clique-cover scheme. Finally, with the help of some examples, we show that the ICC scheme can outperform all of the existing graph-based schemes and the composite-coding scheme.

### A. The ICC scheme vs. the partial-clique-cover scheme

Considering a digraph $D'$ as a $\kappa(D')$-partial clique, we re-express $D'$ in terms of savings in the following proposition.

*Proposition 6:* For a $\kappa(D')$-partial clique, the partial-clique-cover scheme provides savings equal to the minimum out-degree of $D'$, i.e., $S(D') = \delta^+(D')$.

*Proof:* Let the total number of vertices in $D'$ be $N'$. From the definition of partial cliques, i.e., Definition 11, we have

$$\delta^+(D') = N' - (\kappa(D') + 1) = N' - \ell_{\mathsf{PC}}(D') = S(D'). \tag{10}$$

∎

Some partial cliques can be further partitioned into smaller partial cliques such that the sum of the partial-clique numbers of its partitioned partial cliques equals that of the original partial clique. This can cause ambiguity while considering partial cliques during our proofs, so we define the notion of minimal partial clique as follows:

*Definition 17 (Minimal partial clique):* A partial clique $D'$ is a minimal partial clique if the savings obtained from $D'$ (i.e., $\delta^+(D')$ from (10)) is always greater than the sum of the savings in each further partitioned sub-digraphs of $D'$ (over all partitions in $D'$), i.e.,

$$\delta^+(D') > \max_{D'_1, D'_2, \ldots, D'_M} \{\delta^+(D'_1) + \delta^+(D'_2), \ldots + \delta^+(D'_M)\},$$

where $M$ is some integer greater than one, and the maximum is over all partitions $D'_1, D'_2, \ldots, D'_M$ of the minimal partial clique $D'$.

*Theorem 4:* For a class of digraphs, where each digraph $D$ gets partitioned into minimal partial cliques $D'_1, D'_2, \cdots, D'_M$, for some positive integer $m$ such that

(i) for all $i \in \{1, 2, \cdots, M\}$, $\delta^+(D'_i) \in \{0, 1, 2, |V(D'_i)| - 1\}$ (we have four members in this set),

(ii) $V(D'_i) \cap V(D'_j) = \emptyset$ for all $i \neq j$,

(iii) $V(D) = \bigcup_{i=1}^{M} V(D'_i)$, and

(iv) $\ell_{\mathsf{PC}}(D) = \sum_{i=1}^{M} \ell_{\mathsf{PC}}(D'_i)$,

then the ICC scheme performs at least as well as the partial-clique-cover scheme, i.e., $\ell_{\mathsf{ICC}}(D) \leq \ell_{\mathsf{PC}}(D)$.

*Proof:* Refer to Appendix E for the proof. ∎

*Corollary 2:* For any $D$ in which out-degree $d_D^+(i) \leq 2$ for all $i \in V(D)$, we have $\ell_{\mathsf{ICC}}(D) \leq \ell_{\mathsf{PC}}(D)$.

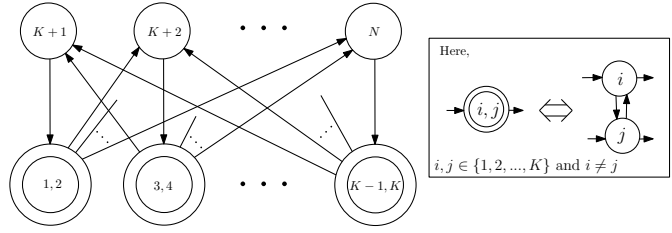

Fig. 5. A digraph belonging to the Class A.

*Proof:* For the given $D$, the minimal partial cliques that partition it to obtain the partial-clique number all have the minimum out-degree belonging to $\{0, 1, 2\}$ because $d_D^+(i) \leq 2$. It follows from Theorem 4 that $\ell_{\mathsf{ICC}}(D) \leq \ell_{\mathsf{PC}}(D)$. ∎

For all digraphs that we have analyzed (including digraphs not considered in Theorem 4), we have observed $\ell_{\mathsf{ICC}}(D) \leq \ell_{\mathsf{PC}}(D)$. For example, the digraph in Fig. 15b has $\ell_{\mathsf{ICC}}(D) = \ell_{\mathsf{PC}}(D)$, and the digraphs in Fig. 1a and Fig. 15a have $\ell_{\mathsf{ICC}}(D)$ strictly less than $\ell_{\mathsf{PC}}(D)$. We conjecture that this holds in general.

*Conjecture 2:* For any digraph $D$, the $\mathsf{ICC}$ scheme performs at least as well as the partial-clique-cover scheme, i.e., $\ell_{\mathsf{ICC}}(D) \leq \ell_{\mathsf{PC}}(D)$.

*1) Additive Gap between the $\mathsf{ICC}$ scheme and the partial-clique-cover scheme:* For an even integer $K \geq 2$, consider a digraph $D$ with $N = \frac{3K}{2}$ vertices. Now we define a class of digraphs, we call *Class A*, as follows: For each $i \in \{1, 2, \ldots, \frac{K}{2}\}$,

(i) vertex $K + i$ has out-degree $d_D^+(K + i) = 2$, and its out-going arcs go to vertices $2i - 1$ and $2i$, and

(ii) vertices $2i - 1$ and $2i$ have $d_D^+(2i - 1) = d_D^+(2i) = \frac{K}{2}$ in such a way that there are arcs in both directions between $2i - 1$ and $2i$, and each of the remaining $\frac{K}{2} - 1$ out-going arcs of $2i - 1$ and $2i$ goes to each vertex in $\{K + 1, K + 2, \ldots, 3K/2\} \setminus \{K + i\}$ (this vertex set has $\frac{K}{2} - 1$ number of vertices in total).

Without loss of generality, a Class A digraph $D$ is shown in Fig. 5.

*Proposition 7:* For any digraph $D$ of Class A, we have $\ell_{\mathsf{PC}}(D) - \ell_{\mathsf{ICC}}(D) = \frac{N}{3} - 1$.

*Proof:* In a digraph $D$ of the Class A (shown in Fig. 5), minimal partial cliques with vertex sets $\{1, 2\}, \{3, 4\}, \ldots, \{K - 1, K\}, \{K + 1\}, \{K + 2\}, \ldots, \{N\}$ provides $\ell_{\mathsf{PC}}(D) = K$ (from Proposition 8 in Appendix F). The digraph $D$ is also an $\mathsf{IC}$ structure with the inner-vertex set $V_I = \{1, 2, \ldots, K\}$ providing $\ell_{\mathsf{ICC}}(D) = \frac{K}{2} + 1$ (from Proposition 8 in Appendix F). Thus

$\ell_{\mathsf{PC}}(D) - \ell_{\mathsf{ICC}}(D) = \frac{K}{2} - 1 = \frac{N}{3} - 1.$ ∎

### B. The ICC scheme vs. the fractional-local-chromatic-number scheme

*Theorem 5:* Consider a $K$-IC structure $D_K$ with $|V(D_K)| = N'$. If

1) the minimum out-degree of $D_K$, $\delta^+(D_K) \leq K - 1$, and

2) $D_K$ has no bi-directional arcs,

then the ICC scheme performs at least as well as the fractional-local-chromatic-number scheme, i.e., $\ell_{\mathsf{ICC}}(D_K) \leq \ell_{\mathsf{FLCN}}(D_K)$.

Before presenting the proof, we start with the definition of the local chromatic number, and a necessary lemma.

*Definition 18 (Local chromatic number [15]):* The local chromatic number $\chi_\ell(D)$ of a digraph $D$ is

$$\chi_\ell(D) := \min_c \max_{i \in V(D)} |\{c(u) : u \in N_D^+(i)\}| + 1,$$

where the minimum is taken over all proper colorings $c$ of $D$, and $c(u)$ is the color of vertex $u$ according to the coloring $c$.

We first prove the following lemma before proving the theorem.

*Lemma 1:* For a digraph $D$ with no bi-directional arcs, the local chromatic number of its complement, i.e., $\bar{D}$, is equal to the difference between the total number of vertices (which is $N$) and the minimum out-degree of the digraph $D$, i.e., $\chi_\ell(\bar{D}) = N - \delta^+(D)$.

*Proof:* Here the underlying undirected graph of $\bar{D}$ is a complete graph, so any proper coloring in $\bar{D}$ yields every vertex with a unique color. Thus the most colorful out-neighborhood belongs to the vertex with the maximum out-degree. For a vertex $v$ in $D$ ($v$ is also in $\bar{D}$), we have

$$\chi_\ell(\bar{D}) = \max_v \ d_{\bar{D}}^+(v) + 1. \tag{11}$$

We know that the out-degree of a vertex $v$ in $D$ and $\bar{D}$ is related by the following equation:

$$d_{\bar{D}}^+(v) = N - 1 - d_D^+(v). \tag{12}$$

From (12), if the vertex $v$ has the minimum out-degree in $D$, then it has the maximum out-degree in $\bar{D}$ and vice versa, so

$$\max_v \ d_{\bar{D}}^+(v) = N - 1 - \min_v \ d_D^+(v) = N - 1 - \delta^+(D). \tag{13}$$

From (11) and (13), we get

$$\chi_\ell(\bar{D}) = N - \delta^+(D). \tag{14}$$

∎

*Proof of Theorem 5:* For a digraph $D$, the length of the index code from the local chromatic number is $\ell_{\mathsf{LCN}}(D) = \chi_\ell(\bar{D})$ [15]. Now for $D_K$ that has no bi-directional arc (given), from Lemma 1 we get,

$$\ell_{\mathsf{LCN}}(D_K) = \chi_\ell(\bar{D}_K) = N' - \delta^+(D_K). \tag{15}$$

Here the underlying undirected graph of $\bar{D}_K$, i.e., $U_{\bar{D}_K}$ is a complete graph, so the fractional local chromatic number $(\chi_f(\bar{D}_K))$ and the local chromatic number $(\chi_\ell(\bar{D}_K))$ is the same, i.e., $\chi_f(\bar{D}_K) = \chi_\ell(\bar{D}_K)$. Hence,

$$\ell_{\mathsf{LCN}}(D_K) = \ell_{\mathsf{FLCN}}(D_K) = N' - \delta^+(D_K). \tag{16}$$

Now considering (7) and (16), we get $\ell_{\mathsf{ICC}}(D_K) \le \ell_{\mathsf{FLCN}}(D_K)$ since $K - 1 \ge \delta^+(D_K)$. ∎

As an example, we construct a class of digraphs satisfying Theorem 5.

*Example 4:* Consider a class of digraphs where each digraph $D$ has an even number of vertices $N$, and let $K = \frac{N}{2}$. Furthermore, vertices in $D$ can be grouped into two sets (without loss of generality, let them be $\{1, 2, \ldots, K\}$ and $\{K + 1, K + 2, \ldots, N\}$) such that for each $i = \{1, 2, \ldots, K\}$, vertex $K + i$ has an arc to vertex $i$, and vertex $i$ has arcs to all $\{K + 1, K + 2, \ldots, N\} \setminus \{K + i\}$. For the digraph $D$, $\delta^+(D) = 1$, and it has no bidirectional arcs. So, $\ell_{\mathsf{FLCN}}(D) = N - \delta^+(D) = N - 1$. One can prove that $D$ is also a $K$-$\mathsf{IC}$ structure with $V_I = \{1, 2, \ldots, K\}$. Now we get $\ell_{\mathsf{ICC}}(D) = N - K + 1 = \frac{N}{2} + 1$. The additive gap, $\ell_{\mathsf{FLCN}}(D) - \ell_{\mathsf{ICC}}(D) = \frac{N-4}{2}$, and the gap grows linearly with $N$. Fig. 3a depicts an example digraph belonging to this class with $K = 3$.

## C. The $\mathsf{ICC}$ scheme can outperform existing schemes including the composite-coding scheme

As an example, we present one $\mathsf{IC}$ structure satisfying case 1 of Theorem 3. The structure is depicted in Fig. 6. The structure is a 4-$\mathsf{IC}$, and it is denoted $D_4$. An index code from the $\mathsf{ICC}$ scheme is $\{x_1 \oplus x_2 \oplus x_3 \oplus x_4, \ x_5 \oplus x_2 \oplus x_3, \ x_6 \oplus x_3 \oplus x_4\}$, which is of length $\ell_{\mathsf{ICC}}(D_4) = 3$. From Theorem 3, $\ell_{\mathsf{ICC}}(D_4) = \beta(D_4) = \beta_t(D_4) = 3$. For $D_4$, the index codelength by existing schemes are shown in Table II (these index codelengths are verified by exhaustive search). The index codelengths provided by the aforementioned existing schemes are strictly greater than $\ell_{\mathsf{ICC}}(D_4)$.

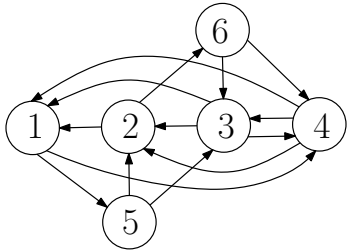

Fig. 6. A 4-IC structure with $V_1 = \{1, 2, 3, 4\}$.

TABLE II

INDEX CODELENGTHS FOR THE DIGRAPH $D_3$ IN FIG. 6 FROM EXISTING SCHEMES

| Schemes | Index codelength |
|---|---|
| Clique cover [13] | 5 |
| Fractional-clique cover [7] | 5 |
| Cycle cover [17], [18] | 4 |
| Fractional-cycle cover [17] | 4 |
| Partial-clique cover [13] | 4 |
| Fractional-partial-clique cover [14] | 4 |
| Local chromatic number [15] | 4 |
| Fractional Local chromatic number [15] | 3.5 |
| Fractional-partitioned local chromatic number [16] | 3.5 |
| Composite coding [9] | 3.5 |

## VIII. EXTENSIONS

In this section, firstly, we extend the ICC scheme using time-sharing to code over overlapping IC structures in a digraph, and to obtain vector linear index codes. Secondly, we extend the definition of the IC structure in such a way that we can extend the ICC scheme to code on extended IC structures.

### A. Fractional ICC scheme

Let $D_s = (V(D_s), \{(i,j) \in A(D) : i, j \in V(D_s)\})$ denote a sub-digraph induced by a subset $s$ of the vertices in $D = (V(D), A(D))$. We define a function

$$\phi(D_s) = \begin{cases} |V(D_s)| - K_s + 1, & \text{if } D_s \text{ is a } K_s\text{-IC structure,} \\ |V(D_s)|, & \text{otherwise.} \end{cases}$$

The index codelength[3] from the fractional ICC scheme is represented as $\ell_{\mathsf{FICC}}(D)$, and given by the following linear program:

$$\text{minimize} \sum_{s \in S} f_s \cdot \phi(D_s)$$

$$\text{subject to} \sum_{s \in S: j \in V(D_s)} f_s \geq 1, \quad \text{for each } j \in V(D), \tag{17}$$

$$f_s \in [0, 1], \ s \in S.$$

Here $S$ is the power set of $V(D)$. In the fractional ICC scheme, each sub-digraph induced by the subset $s$ in $S$ is assigned a weight $[0, 1]$ such that the total weight of each message over all of the subsets it belongs to is at least one. In this scheme, $\ell_{\mathsf{FICC}}(D)$ is the minimum sum of weights. The ICC scheme is a special case of the fractional ICC scheme where $f_s \in \{0, 1\}$, so $\beta(D) \leq \ell_{\mathsf{FICC}}(D) \leq \ell_{\mathsf{ICC}}(D)$.

### B. Extension of the IC structure

We start with an example that provides an insight to the extension of the IC structure. Consider a digraph $D$ that has three cliques $\rho_1$, $\rho_2$ and $\rho_3$ each of size two. Let $V(\rho_1) = \{1, 2\}, V(\rho_2) = \{3, 4\}$ and $V(\rho_3) = \{5, 6\}$ be vertex sets of those three cliques respectively. Furthermore, for the clique pair $(\rho_1, \rho_2)$, all vertices in $V(\rho_1)$ have out-going arcs to all vertices in $V(\rho_2)$, and the result follows similarly for clique pairs $(\rho_2, \rho_3)$ and $(\rho_3, \rho_1)$. This digraph is depicted in Fig. 7a. One can verify that $D$ is a 4-IC structure with an inner vertex set $V_{\mathsf{I}} = \{1, 2, 3, 4\}$. We cannot get a 5-IC structure in $D$. Suppose that we pick $V_{\mathsf{I}} = \{1, 2, 3, 4, 5\}$, there is no path from vertices 1 and 2 to vertex 5 without passing through the inner vertex 3 or 4. By symmetry, choosing any 5 vertices as an inner vertex set will have the same issue.

Now the ICC scheme gives an index code $\{x_1 \oplus x_2 \oplus x_3 \oplus x_4, \ x_5 \oplus x_1 \oplus x_2, \ x_6\}$ of length $\ell_{\mathsf{ICC}}(D) = 3$. However, from the coding point of view of an IC structure, vertices 5 and 6 need not be separated during encoding because they have the same arc sets $N_D^+(5) \setminus \{6\} = N_D^+(6) \setminus \{5\}$ and $N_D^-(5) \setminus \{6\} = N_D^-(6) \setminus \{5\}$, and have arcs to each other. This means we can get another index code by removing $x_6$, and replacing $x_5$ with $x_5 \oplus x_6$, i.e., $\{x_1 \oplus x_2 \oplus x_3 \oplus x_4, \ x_5 \oplus x_6 \oplus x_1 \oplus x_2\}$ of length two. Here due to the special connectivity of clique $\{5, 6\}$, we have treated it as a single vertex, and used $x_5 \oplus x_6$ in the code construction. In light of this, we will now extend the

---

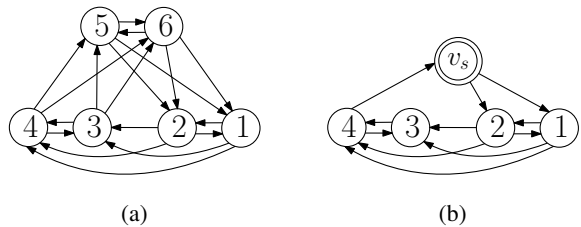

(a)                                    (b)

Fig. 7.   (a) The digraph with three cliques each of order two, and all vertices of a clique have out-going arcs to all vertices of another clique such that the three cliques are connected in a cyclic pattern as shown (from the ICC scheme, the digraph has an index code $\{x_1 \oplus x_2 \oplus x_3 \oplus x_4,\ x_5 \oplus x_1 \oplus x_2,\ x_6\}$ of length $\ell_{\mathsf{ICC}}(D) = 3$), and (b) the same digraph, but with a super-vertex notation (the super-vertex, denoted by $v_s$, is formed by vertices 5 and 6, and the digraph has an index code $\{x_1 \oplus x_2 \oplus x_3 \oplus x_4,\ x_5 \oplus x_6 \oplus x_1 \oplus x_2\}$ of length two).

definition of an IC structure to capture cliques with such special configurations. To achieve this we define a term called a super-vertex.

*Definition 19 (Super-vertex):* In a digraph $D$, let $V_s$ be a vertex set where (i) all vertices in $V_s$ have arcs to each other, i.e., they form a clique, and (ii) every vertex $i \in V_s$ has the same $N_D^+(i) \setminus V_s$ and the same $N_D^-(i) \setminus V_s$. Such a group of vertices (all $i \in V_s$) is called a super-vertex and denoted as $v_s$.

Now we define extended IC structures and an index-coding scheme for them.

*Definition 20 (Extended IC structure):* The extended IC (EIC) structure is defined as an IC structure that allows super-vertices in its non-inner vertex set.

*Definition 21 (Extended ICC scheme):* For any digraph $D$, the extended ICC (EICC) scheme finds a set of disjoint EIC structures covering $D$. It then codes each of these EIC structures using the code construction described in the following:

- Each super-vertex (non-inner vertices) is treated as a single vertex during the construction and the encoding process of the EIC structure.

- We consider the message requested by the super-vertex to be the XOR of all messages requested by the vertices forming the super-vertex.

- Each of these EIC structures are treated as an IC structure, and an index code is constructed using the ICC scheme.

Along with super-vertices, and taking their definition into account, one can prove the validity of the code constructed by the EICC scheme similar to the proof of Proposition 4. Denote the length of the index code produced by the EICC scheme by $\ell_{\mathsf{EICC}}(D)$.

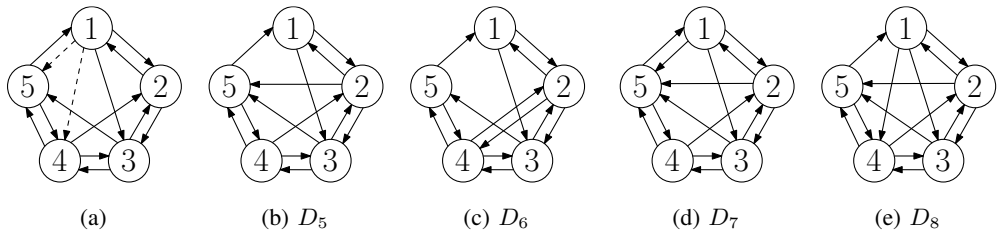

(a)   (b) $D_5$   (c) $D_6$   (d) $D_7$   (e) $D_8$

Fig. 8. (a) We form four digraphs $D_1, D_2, D_3,$ and $D_4$ by removing any number (zero to two inclusive) of dashed arcs from the digraph (this gives four non-isomorphic digraphs), (b) $D_5$, (c) $D_6$, (d) $D_7$, and (e) $D_8$. All of the digraphs ($D_1$ to $D_8$) have an index code $\{x_1 \oplus x_2 \oplus x_3, \; x_1 \oplus x_4 \oplus x_5\}$ of length two. Furthermore, two is the optimal codelength for the digraphs. The ICC scheme (including fractional and extended versions) achieves the minimum of the index codelength of (i) $\frac{7}{3}$ for $D_6$, and (ii) $\frac{5}{2}$ for the remaining digraphs $D_1, D_2, D_3, D_4, D_5, D_7,$ and $D_8$.

*Theorem 6:* For a digraph $D$, the index codelength obtained from the EICC scheme is a better upper bound to the optimal broadcast rate than the codelength obtained from the ICC scheme, i.e., $\beta(D) \leq \beta_t(D) \leq \ell_{\mathsf{EICC}}(D) \leq \ell_{\mathsf{ICC}}(D)$.

*Proof:* It follows from the definition of EIC structures that include IC structures as special cases. ∎

*Remark 9:* Referring to the work by Ong [24], which proved that linear index codes are optimal for all unicast index-coding instances up to and including five receivers, we find that the ICC scheme (including fractional and extended versions) provides the minimum index codelength for all message alphabet sizes for all unicast index-coding instances up to five receivers (9846 non-isomorphic problems) except 8 problems (modeled by 8 non-isomorphic digraphs). The eight non-isomorphic digraphs are illustrated in Fig. 8.

## IX. CONCLUSION

Graph-based approaches have been shown to be useful for index coding, in which cycles play an important role. Prior to this work, cycles and cliques (including the timeshared version) were exploited to construct index codes. In this work, we attempted to extend the role of cycles in index coding. We took a step further and showed the benefits of coding on interlinked-cycle structures (one form of overlapping cycles). Our proposed scheme generalizes coding on cycles and cliques. By identifying a useful interlinked-cycle structure, we were able to characterize a class of infinitely many graphs where scalar linear index codes are optimal. For some classes of digraphs, we proved that the ICC scheme performs at least as well as some existing schemes such as the partial-clique-cover scheme and the fractional-local-chromatic-number scheme. Furthermore, for a class of digraphs, we proved that the partial-clique-cover scheme and the ICC scheme have linearly-growing additive gap in index codelength with the number of vertices in the digraphs. We proved a similar result for the fractional-local-chromatic-number scheme and the ICC scheme for another class of digraphs. We extended the ICC scheme, to allow time-sharing over all possible IC structures in digraphs. We also extended the IC structure to allow super vertices as its non-inner vertices. However, it remains an open problem to identify cycles overlapping in other useful ways.

## Appendix A

### Proof of Proposition 4

A $K$-IC structure $D_K$ has some properties captured in the following lemmas, which will be used to prove Proposition 4. Here we consider $T_i$ and $T_j$ as any two distinct directed rooted trees present in $D_K$ with the root vertices $i$ and $j$ respectively.

*Lemma 2:* For any vertex $v \in V(T_i) \cap V(T_j)$, and $v \notin V_I$, the set of leaf vertices that fan out from the common vertex $v$ in each tree is a subset of $V_I \setminus \{i, j\}$.

*Proof:* In a tree $T_i$ (see Fig. 9), for any vertex $v \in V(T_i)$ and $v \notin V_I$, let $L_{T_i}(v)$ be a set of leaf vertices that fan out from vertex $v$. If vertex $j \in L_{T_i}(v)$, then there exists a path from $v$ to $j$ in $T_i$. However, in $T_j$, there is a path from $j$ to $v$. Thus in the sub-digraph[4] $D_K$, we obtain a path from $v$ to $j$ (via $T_i$) and vice versa (via $T_j$). As a result, a cycle including non-inner vertices and only one inner vertex (i.e., $j$) exists. This cycle is an I-cycle, and condition 1 (i.e., no I-cycle) for $D_K$ is violated. Hence, $j \notin L_{T_i}(v)$. In other words, $L_{T_i}(v) \subseteq V_I \setminus \{i, j\}$. Similarly, $L_{T_j}(v) \subseteq V_I \setminus \{i, j\}$. ∎

*Lemma 3:* For any vertex $v \in V(T_i) \cap V(T_j)$, and $v \notin V_I$, the out-neighborhood of vertex $v$ is same in both trees, i.e., $N_{T_i}^+(v) = N_{T_j}^+(v)$.

*Proof:* Here the proof is done by contradiction. Let us suppose that $N_{T_i}^+(v) \neq N_{T_j}^+(v)$.

For this proof we refer to Fig. 9. This proof has two parts. In the first part, we prove that $L_{T_i}(v) = L_{T_j}(v)$, and then prove that $N_{T_i}^+(v) = N_{T_j}^+(v)$ in the second part.

(Part 1) Suppose that $L_{T_i}(v) \neq L_{T_j}(v)$. From Lemma 2, $L_{T_i}(v)$ is a subset of $V_I \setminus \{i, j\}$. Now pick a vertex $c$ that belongs to $V_I \setminus \{i, j\}$ such that $c \in L_{T_i}(v)$ but $c \notin L_{T_j}(v)$ (such $c$ exists since we suppose that $L_{T_i}(v) \neq L_{T_j}(v)$, and we swap the indices $i$ and $j$ if $L_{T_i}(v) \subset L_{T_j}(v)$). In tree $T_i$, there exists a directed path from vertex $i$, which includes $v$ to the leaf vertex $c$. Let this path be $P_{i \to c}(T_i)$. Similarly, in tree $T_j$, there exists a directed path from vertex $j$, which doesn't include $v$ (since $c \notin L_{T_j}(v)$), and ends at the leaf vertex $c$. Let this path be $P_{j \to c}(T_j)$. In the digraph $D_K$, we can also obtain a directed path from $j$ which passes through $v$ (via $T_j$), and ends at the leaf vertex $c$ (via $T_i$). Let this path be $P_{j \to c}(D_K)$. The paths $P_{j \to c}(T_j)$ and $P_{j \to c}(D_K)$ are different, which indicates the existence of multiple I-paths from $j$ to $c$ in $D_K$, this violates the condition 2 for $D_K$. Consequently, $L_{T_i}(v) = L_{T_j}(v)$.

---

[4]As $D_K = \bigcup_{\forall i \in V_I} T_i$, a path present in any $T_i$ also present in $D_K$.

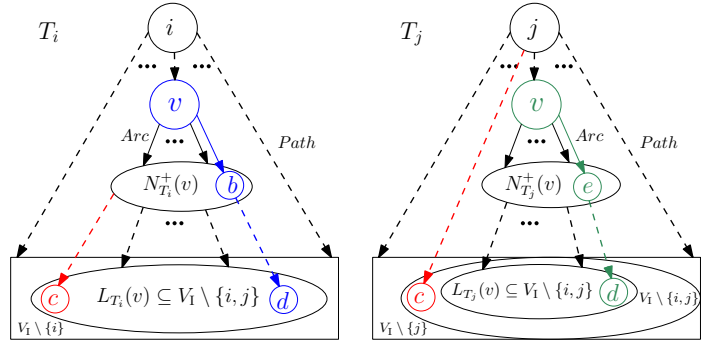

Fig. 9. Outline of the directed rooted trees $T_i$ and $T_j$ with root vertices $i$ and $j$ respectively, and a non-inner vertex $v$ in common. Here we have used solid arrow to indicate an arc, and dashed arrow to indicate a path.

(Part 2) Now we pick a vertex $b$ such that, without loss of generality, $b \in N_{T_i}^+(v)$ but $b \notin N_{T_j}^+(v)$ (such $b$ exists since we assumed that $N_{T_i}^+(v) \neq N_{T_j}^+(v)$, and we swap the indices $i$ and $j$ if $N_{T_i}^+(v) \subset N_{T_j}^+(v)$). Furthermore, we have two cases for $b$, which are (case 1) $b \in L_{T_i}(v)$, and (case 2) $b \notin L_{T_i}(v)$. Case 1 is addressed in the first part of this proof. On the other hand, for case 2, we pick a leaf vertex $d \in L_{T_i}(b)$ such that there exists a path (see Fig. 9) that starts from $v$ followed by $b$, and ends at $d$, i.e., $\langle v, b, \ldots, d \rangle$ exists in $T_i$. A path $\langle j, \ldots, v \rangle$ exists in $T_j$. Thus a path $\langle j, \ldots, v, b, \ldots, d \rangle$ exists in $D_K$. From the first part of the proof, we have $L_{T_i}(v) = L_{T_j}(v)$, so $d \in L_{T_j}(v)$. Now in $T_j$, there exists a path from $j$ to $d$, which includes $v$ followed by a vertex $e$ such that $e \in N_{T_j}^+(v)$ and $e \neq b$ (as $b \notin N_{T_j}^+(v)$), and the path ends at $d$, i.e., $\langle j, \ldots, v, e, \ldots, d \rangle$ which is different from $\langle j, \ldots, v, b, \ldots, d \rangle$. Note that in trees $T_i$ and $T_j$, only the root and the leaf vertices are from $V_I$, so multiple I-paths are observed at $d$ from $j$. This violates condition 2 for $D_K$. Consequently, $N_{T_i}^+(v) = N_{T_j}^+(v)$. ∎

*Lemma 4:* If a vertex $v \in V(T_i)$ such that $v \notin V_I$, then its out-neighborhood is the same in $T_i$ and in $D_K$, i.e., $N_{T_i}^+(v) = N_{D_K}^+(v)$.

*Proof:* For any $v \in V(T_i)$ from Lemma 3, $N_{T_i}^+(v) = N_{T_j}^+(v)$ for all $\{j : v \in T_j\}$. Since $D_K = \bigcup_{\forall i \in V_I} T_i$, vertex $v$ must have the same out-neighborhood in $D_K$ as well. ∎

*Proof of Proposition 4:* From (6), all $j \in \{K + 1, K + 2, \ldots, N'\}$ which are non-inner vertices, can decode their requested messages. This is because the coded symbol $w_j$ is the bitwise XOR of the messages requested by $j$ and its all out-neighborhood vertices, and any $j$ has messages requested by all of its out-neighborhood vertices in its side information.

For an inner vertex $i$, rather than analyzing the sub-digraph $D_K$, we will analyze its tree $T_i$, and show that it can decode its message from the relevant symbols in $W$. We are able to consider only the tree $T_i$ due to the Lemma 4. Now let us take any tree $T_i$. Assume that it has a height $H$ where $1 \leq H \leq (N' - K + 1)$. The vertices in $T_i$ are at various depths, i.e., $\{0, 1, 2, \ldots, H\}$ from the root vertex $i$. The root vertex $i$ has depth zero, and any vertex at depth equal to the height of the tree is a leaf vertex.

Firstly, in $T_i$, we compute the bitwise XOR among coded symbols of all non-leaf vertices at depth greater than zero, i.e., $Z_i \triangleq \bigoplus_{j \in V(T_i) \setminus V_I} w_j$. However, in $T_i$, the message requested by a non-leaf vertex, say $p$, at a depth strictly greater than one, appears exactly twice in $\{w_j : j \in V(T_i) \setminus V_I\}$;

  i) once in $w_k$, where $k$ is parent of $p$ in tree $T_i$, and

  ii) once in $w_p$. Refer to (20) for mathematical details.

Thus they cancel out each other while computing $Z_i$ in the tree $T_i$. Hence, in the tree $T_i$, the resultant expression is the bitwise XOR of

  i) messages requested by all non-leaf vertices at depth one, and

  ii) messages requested by all leaf vertices at depth strictly greater than one.

Refer to (21) for mathematical details.

Secondly, in $T_i$, we compute $w_I \oplus Z_i$ (refer to (22) for mathematical details) which yields the bitwise XOR of

  i) the messages requested by all non-leaf vertices at depth one, which are out-neighbors of $i$,

  ii) the messages requested by all leaf vertices at depth one, which are also out-neighbors of $i$, and

  iii) the message requested by $i$, i.e., $x_i$.

This is because the message requested by each leaf vertex at depth strictly greater than one in the tree $T_i$ is present in both the resultant terms of $Z_i$ and in $w_I$, thereby canceling out itself in $w_I \oplus Z_i$. Hence, $w_I \oplus Z_i$ yields the bitwise XOR of $x_i$ and $\{x_j : j \in N_{D_K}^+(i)\}$. As $i$ knows all $\{x_j : j \in N_{D_K}^+(i)\}$ as side-information, any inner vertex $i$ can decode its required message from $w_I \oplus Z_i$.

The mathematical computations of $Z_i$ and $w_I \oplus Z_i$ in the tree $T_i$ are as follows:

*1) Computing $Z_i$:*

$$
\begin{aligned}
Z_i &= \bigoplus_{j \in V(T_i) \setminus V_I} w_j \\
&= \bigoplus_{j \in V(T_i) \setminus V_I} \left( x_j \oplus \bigoplus_{q \in N_{D_K}^+(j)} x_q \right) \\
&= \bigoplus_{j \in V(T_i) \setminus V_I} \left( x_j \oplus \bigoplus_{q \in N_{T_i}^+(j)} x_q \right) \\
&= \bigoplus_{j \in V(T_i) \setminus V_I} \left( x_j \oplus \bigoplus_{q \in N_{T_i}^+(j) \setminus V_I} x_q \oplus \bigoplus_{q \in N_{T_i}^+(j) \cap V_I} x_q \right) \\
&= X_{V(T_i)} \oplus X'_{V(T_i)}.
\end{aligned}
\tag{18}
$$

Where,

$$
X_{V(T_i)} \triangleq \bigoplus_{j \in V(T_i) \setminus V_I} \left( x_j \oplus \bigoplus_{q \in N_{T_i}^+(j) \setminus V_I} x_q \right), \text{ and}
$$

$$
X'_{V(T_i)} \triangleq \bigoplus_{j \in V(T_i) \setminus V_I} \left( \bigoplus_{q \in N_{T_i}^+(j) \cap V_I} x_q \right) = \bigoplus_{\substack{q : q \in V_I \setminus \{i\} \\ \& \ q \notin N_{T_i}^+(i)}} x_q.
\tag{19}
$$

Here (19) is obtained because each $q \in V_I \setminus \{i\}$ has only one parent in $T_i$, and we exclude all $q \in V_I \setminus \{i\}$ whose parent is $i$, and $X'_{V(T_i)}$ is bitwise XOR of messages requested by all of the leaf vertices not in the out-neighborhood of $i$. If we expand $X_{V(T_i)}$ as per the group of vertices according to their depth, we get

$$
\begin{aligned}
X_{V(T_i)} \\
= \bigoplus_{j_1 \in N_{T_i}^+(i) \setminus V_I} & \left[ \left( x_{j_1} \oplus \bigoplus_{q \in N_{T_i}^+(j_1) \setminus V_I} x_q \right) \oplus \right.\\
& \bigoplus_{j_2 \in N_{T_i}^+(j_1) \setminus V_I} \left[ \left( x_{j_2} \oplus \bigoplus_{q \in N_{T_i}^+(j_2) \setminus V_I} x_q \right) \oplus \ldots \oplus \right.\\
& \bigoplus_{\substack{j_{H-2} \in \\ N_{T_i}^+(j_{H-3}) \setminus V_I}} \left[ \left( x_{j_{H-2}} \oplus \bigoplus_{q \in N_{T_i}^+(j_{H-2}) \setminus V_I} x_q \right) \oplus \right.\\
& \left. \left. \left. \bigoplus_{\substack{j_{H-1} \in \\ N_{T_i}^+(j_{H-2}) \setminus V_I}} \left[ \left( x_{j_{H-1}} \oplus \underbrace{\bigoplus_{q \in N_{T_i}^+(j_{H-1}) \setminus V_I} x_q}_{=\emptyset} \right) \right] \ldots \right] \right] \right] \\
= \bigoplus_{j \in N_{T_i}^+(i) \setminus V_I} & x_j.
\end{aligned}
\tag{20}
$$

Note that the intermediate terms in $X_{V(T_i)}$ cancel out (we have used the same color to indicate the terms that cancel each other). Now substituting $X_{V(T_i)}$ of (20) and $X'_{V(T_i)}$ of (19) in (18), we get

$$
Z_i = \bigoplus_{k \in N_{T_i}^+(i) \setminus V_I} x_k \oplus \left( \bigoplus_{\substack{q : q \in V_I \setminus \{i\} \\ \& \ q \notin N_{T_i}^+(i)}} x_q \right).
\tag{21}
$$

*2) Computing $w_I \oplus Z_i$:*

$$w_I \oplus Z_i = w_I \oplus X_{V(T_i)} \oplus X'_{V(T_i)}$$

$$= x_i \oplus \bigoplus_{j \in V_I \setminus \{i\}} x_j \oplus \bigoplus_{k \in N_{T_i}^+(i) \setminus V_I} x_k \oplus \bigoplus_{\substack{q: q \in V_I \setminus \{i\} \\ \& \ q \notin N_{T_i}^+(i)}} x_q$$

$$= x_i \oplus \left( \bigoplus_{\substack{j: j \in V_I \setminus \{i\} \\ \& \ j \in N_{T_i}^+(i)}} x_j \oplus \bigoplus_{\substack{j: j \in V_I \setminus \{i\} \\ \& \ j \notin N_{T_i}^+(i)}} x_j \right) \oplus$$

$$\bigoplus_{k \in N_{T_i}^+(i) \setminus V_I} x_k \oplus \left( \bigoplus_{\substack{q: q \in V_I \setminus \{i\} \\ \& \ q \notin N_{T_i}^+(i)}} x_q \right)$$

$$= x_i \oplus \left( \bigoplus_{\substack{j: j \in V_I \setminus \{i\} \\ \& \ j \in N_{T_i}^+(i)}} x_j \oplus \bigoplus_{k \in N_{T_i}^+(i) \setminus V_I} x_k \right). \tag{22}$$

∎

## Appendix B

### An algorithm for the ICC scheme

For the ICC scheme, we present an algorithm that exhaustively searches all IC structures in a digraph $D$ and computes its index codelength. The overview of the algorithm is as follows: In the first round, we remove all the vertices having in-degree or out-degree equal to zero and the arcs associated with them from $D$, and consider the resulting sub-digraph, denoted $D'$, for further processing. In the second round, we start with a single vertex $i \in V(D')$ (which is considered as a 1-IC structure) and check whether or not, in $D'$, we can find larger IC structures with the inner-vertex set $V_I = \{i, j\}$ for each $j \in V(D')$, $j \neq i$. This checking and finding task is done by invoking a proposed function, which we call $\text{CheckInnerVertices}(V_I', D_K, v', D', \mathbb{P})$, where $D_K$ is an IC structure, $V_I'$ is the inner-vertex set of $D_K$, $v'$ is the vertex that need to be checked, and $\mathbb{P}$ is the set of all possible simple paths in $D'$ between all pairs of the unique vertices in $V(D')$ (for this round, $V_I' = V_I$, $D_K = (\{i\}, \emptyset)$, $v' = j$, and $D'$ and $\mathbb{P}$ remain the same for all round). The pseudocode of the function is presented as Algorithm 2. In $D'$, there can exist multiple paths between any two vertices, so we can obtain multiple IC structures with the same $V_I$, and the

---

**Algorithm 1:** Enumerate all possible IC structures for a digraph $D$ and compute $\ell_{\mathsf{ICC}}(D)$.

---

**Input:** Digraph $D = (V(D), A(D))$. $//V(D) = \{1, 2, \ldots, N\}$.

**Output:** $\ell_{\mathsf{ICC}}(D)$ and the corresponding IC structures covering $D$.

$V_0 = \{v : v \in V(D) \text{ and } d_D^+(v) = 0, \text{ or } d_D^-(v) = 0\}$;

$V(D') = V(D) \setminus V_0$;

Let a sub-digraph of $D$ induced by the vertex set $V(D')$ be $D'$, where $D' = (V(D'), A(D'))$ and without loss of generality

$V(D') = \{1, 2, \cdots, N'\}$; $N' \geq 1$;

```
//Initialization
```

$\mathbb{I} = \{(\emptyset, \emptyset)\}$; `  //A family of sets collecting all` IC `structures along with their`
`inner-vertex sets.`

For any $i, j \in V(D')$ such that $i \neq j$, let $\mathbb{P}_{i \to j}(D') \triangleq \{P_{i \to j}(D')\}$ be the set of all possible simple paths from $i$ to $j$; `  //All`
`possible simple paths between any two vertices in a digraph can be found by`
`implementing Warshall's algorithm or the modified depth-first search algorithm.`

$\mathbb{P} \triangleq \{\mathbb{P}_{i \to j}(D')\}_{\forall i, j \in V(D'): i \neq j}$;

**foreach** *vertex* $i \in \{1, 2, \ldots, N'\}$ **do**

> $V_{1,i} = \{i\}$;
>
> $\mathbb{D}_{1,i} = \{D_{1,i,1}\} = \{(\{i\}, \emptyset)\}$; `          //`$\mathbb{D}_{a,b}$ `is a family of sets collecting all` IC
> `structures with the same inner vertex set` $V_{a,b}$`, where` $a$ `indicates the total`
> `number of inner vertices and` $b$ `is used for indexing, and` $D_{a,b,j}$ `is an` IC `structure`
> `which is the` $j$`th element of` $\mathbb{D}_{a,b}$`.`
>
> $\mathbb{I} \leftarrow \mathbb{I} \cup \{(D_{1,i,1}, V_{1,i})\}$;

$s = 1$;

$P_s = N'$;

$\mathbb{V} = \emptyset$; `   //`$\mathbb{V}$ `is a family of sets collecting all sets of vertices that have already`
`been checked whether or not they are inner-vertex sets of` IC `structures.`

```
//Iteration
```

**while** $P_s \neq 0$ **do**

> $p = 0$;
>
> **foreach** $V_{s,i}, \ i \in \{1, 2, \ldots, P_s\}$ **do**
>
> > **foreach** *non-empty* $D_{s,i,j} \in \mathbb{D}_{s,i}, \ j \in \{1, 2, \ldots, |\mathbb{D}_{s,i}|\}$ **do**
> >
> > > $V' = V(D') \setminus V(D_{s,i,j})$;
> > >
> > > **if** $V' \neq \emptyset$ **then**
> > >
> > > > **foreach** $k \in V'$ **do**
> > > >
> > > > > **if** $(V_{s,i} \cup \{k\}) \notin \mathbb{V}$ **then**
> > > > >
> > > > > > (check, $\mathbb{D}$) = CheckInnerVertices($V_{s,i}, \ D_{s,i,j}, \ k, \ D', \ \mathbb{P}$);
> > > > > >
> > > > > > **if** check = Yes **then**
> > > > > >
> > > > > > > $p \leftarrow p + 1$;
> > > > > > >
> > > > > > > $V_{s+1,p} = V_{s,i} \cup \{k\}$;
> > > > > > >
> > > > > > > $\mathbb{D}_{s+1,p} = \mathbb{D} \triangleq \{D_{s+1,p,1}, D_{s+1,p,2}, \ldots, D_{s+1,p,M}\}$, where $M = |\mathbb{D}|$;
> > > > > > >
> > > > > > > $\mathbb{I} \leftarrow \mathbb{I} \cup \{(D_{s+1,p,m}, V_{s+1,p})\}_{\forall m \in \{1, 2, \ldots, M\}}$;
> > > > >
> > > > > $\mathbb{V} \leftarrow \mathbb{V} \cup \{(V_{s,i} \cup \{k\})\}$;
>
> $s \leftarrow s + 1$;
>
> $P_s \leftarrow p$;

Let cover($\mathbb{I}$) $\triangleq \{\mathbb{I}' \subseteq \mathbb{I} : \bigcup_{D_{a,b,c} \in \mathbb{I}'} V(D_{a,b,c}) = V(D')\}$, where $a, b, c$ are positive integers providing the label of the chosen IC structures covering $D'$;

$\ell_{\mathsf{ICC}}(D') = \min\limits_{\mathbb{I}' \in \text{ cover}(\mathbb{I})} \sum\limits_{\forall D_{a,b,c} \in \mathbb{I}'} (|V(D_{a,b,c})| - |V_{a,b}| + 1)$, where $\ell_{\mathsf{ICC}}(D')$ is minimized over all $\mathbb{I}'$;

$\ell_{\mathsf{ICC}}(D) = |V_0| + \ell_{\mathsf{ICC}}(D')$;

---

---

**Algorithm 2:** Checks and grows a given IC structure.

---

**Function** CheckInnerVertices($V'_I, D_K, v', D', \mathbb{P}$)

  Without loss of generality, let $V'_I = \{1, 2, \ldots, M\}$, where $M = |V'_I|$;

  For any $i, j \in V(D')$, consider $\mathbb{P}'_{i \to j}(D') \triangleq \mathbb{P}_{i \to j}(D') \setminus \{P_{i \to j}(D') : V(P_{i \to j}(D')) \cap (V'_I \setminus \{i, j\}) \neq \emptyset\}$;

  $s = 1$;

  $\mathbb{D}' = \emptyset$;   `//D' is a family of sets collecting all of the IC structures grown from`
     `the given IC structure` $D_K$.

  **if** $\forall i \in V'_I, \mathbb{P}'_{i \to v'}(D') \neq \emptyset$ *and* $\mathbb{P}'_{v' \to i}(D') \neq \emptyset$ **then**

    **foreach** *unique*

$$[P_{1 \to v'}(D'), \ldots, P_{M \to v'}(D'), \ P_{v' \to 1}(D'),$$
$$\ldots, P_{v' \to M}(D')] \in \mathbb{P}'_{1 \to v'}(D') \times \cdots \times$$
$$\mathbb{P}'_{M \to v'}(D') \times \mathbb{P}'_{v' \to 1}(D') \times \cdots \times \mathbb{P}'_{v' \to M}(D')$$

    **do**

      $D'_s = D_K \cup (P_{1 \to v'}(D') \cup \ldots \cup P_{M \to v'}(D') \cup P_{v' \to 1}(D') \cup \ldots \cup P_{v' \to M}(D'))$;

      $V' = V'_I \cup \{v'\}$;

      `//In` $D'_s$, `we check I-cycles and I-paths`

      **if** $\forall i \in V'$, *every* $P_{i \to i}(D'_s)$ *has* $V(P_{i \to i}(D'_s)) \cap (V' \setminus \{i\}) \neq \emptyset$ **then**

        **if** $\forall i, j \in V'$, $i \neq j$, *there is one and only one path* $P_{i \to j}(D'_s)$ *such that* $V(P_{i \to j}(D'_s)) \cap V' = \{i, j\}$ **then**

          $\mathbb{D}' \leftarrow \mathbb{D}' \cup \{D'_s\}$;

          $s \leftarrow s + 1$;

  **if** $\mathbb{D}' \neq \emptyset$ **then**

    **return** (Yes, $\mathbb{D}'$);

  **else**

    **return** (No, $\mathbb{D}'$);

---

function CheckInnerVertices($V'_I, D_K, v', D', \mathbb{P}$) finds and returns all possible IC structures with the same $V_I$. We carry out the second round operations for all vertices in $V(D')$. In the third round, for each of the IC structures obtained in the second round, we check whether or not we can find a larger IC structure that includes one extra inner-vertex from the vertices not included in the IC structure. This is again done by invoking the function CheckInnerVertices($V'_I, D_K, v', D', \mathbb{P}$). We repeat this process of checking and finding IC structures till we can not find any larger IC structures. This will happen when the IC structure in the earlier round is the largest IC structure possible with the respective vertices. For details, refer to the pseudocode in Algorithm 1. After finding all possible IC structure in $D'$, we compute $\ell_{ICC}(D')$ by minimizing the codelengths provided by IC structures covering $D'$ over all possible IC structures in $D'$. The sum of $\ell_{ICC}(D')$ and the number of vertices removed from $D$ in the first round provides $\ell_{ICC}(D)$.

The algorithm presented here is a brute-force search, and it might be intractable for large digraphs (having large number of vertices) because of large numbers of checks that need to be performed during the algorithm's runtime. We do not consider designing heuristic (but faster) algorithms, as it is not the focus of this work.

## Appendix C
### Proof of Theorem 3

We first prove one lemma that will help to prove the optimality of the ICC scheme.

*Lemma 5:* In an IC structure, any cycle must contain either (i) no inner vertex, or (ii) at least two inner vertices.

*Proof:* It follows directly from the property of an IC structure that a cycle cannot be formed by including only one inner vertex because this type of cycle is an I-cycle. ∎

*Proof of Theorem 3:* We will show that the MAIS lower bound (1) is tight for all $t$. We consider that $D_K$ has $N' \geq 1$ vertices. For any $D_K$, if $K = 1$, then there exists an inner vertex, and there is no arc and vertex except the inner vertex. This is because in $D_{K=1}$, if there exists any other vertex beside the inner vertex, then it must be a non-inner vertex because $K = 1$. The additional vertex can be a non-inner vertex if and only if it is included by an I-path in $D_{K=1}$. If there exists an I-path, then there must be at least two inner vertices in $D_{K=1}$ (which contradicts the given condition $K = 1$). Thus $D_{K=1}$ contains only one vertex, and $\mathsf{MAIS}(D_{K=1}) = 1$. For $K \geq 2$, we have the following:

(Case 1) From Lemma 5, any cycle must include at least two inner vertices, or no inner vertex, thus if we remove $K - 1$ inner vertices, then the digraph $D_K$ becomes acyclic. Thus

$$\mathsf{MAIS}(D_K) \geq N' - K + 1. \tag{23}$$

From Theorem 1, we get

$$\ell_{\mathsf{ICC}}(D_K) = N' - K + 1. \tag{24}$$

It follows from (1), (23) and (24) that $\mathsf{MAIS}(D_K) = N' - K + 1 = \ell_{\mathsf{ICC}}(D_K)$. Thus $\ell_{\mathsf{ICC}}(D_K) = \beta(D_K) = \beta_t(D_K) = N' - K + 1$.

(Case 2) A $D_K$ can be viewed in two ways. The first way is considering the whole $D_K$ as a $K$-IC structure. The second way is considering induced sub-digraphs of $D_K$ which consist of

    a) $M$ disjoint cycles together consisting of a total of $N_A$ ($0 \leq N_A < N' - K$) non-inner vertices (if $N_A = 0$ or 1, then $M = 0$, which is case 1),

b) $M + 1$ disjoint IC structures each with $N_i$ vertices and $K_i$ inner vertices in such a way that $\sum_{i=1}^{M+1} K_i = K$, we consider that each IC structure is also disjoint from all $M$ cycles among non-inner vertices, and

c) total remaining of $N_B = N' - N_A - \sum_{i=1}^{M+1} N_i$ non-inner vertices (which are not included in $M$ cycles, or the $M + 1$ IC structures).

Now we will show that both ways of looking at $D_K$ are equivalent in the sense of the index codelength generated from our proposed scheme, and both equal to $\mathsf{MAIS}(D_K)$. We prefer the second way of viewing $D_K$ for our proof since it is easier to find the MAIS lower bound.

For the partitioned $D_K$ (looking at in the second way), the total number of coded symbols is the summation of the coded symbols for (i) each of the $M$ disjoint cycles (each cycle has saving equal to one), (ii) each of the $M + 1$ disjoint IC structures (each of the IC structures has savings equal to $K - 1$), and (iii) $N_B$ uncoded symbols for the remaining non-inner vertices, i.e.,

$$\ell'_{\mathsf{ICC}}(D_K) = (N_A - M) + \sum_{i=1}^{M+1}(N_i - K_i + 1) + N_B$$

$$= N' - K + 1. \qquad (25)$$

From (24) and (25), $\ell_{\mathsf{ICC}}(D_K) = \ell'_{\mathsf{ICC}}(D_K)$, thus from both perspectives the code length is the same.

Now for $D_K$ (looking at in our second way), if we remove one vertex from each of the $M$ cycles among non-inner vertices ($M$ removal in total), and remove $K_i - 1$ vertices from each of the $M + 1$ IC structures ($\sum_{i=1}^{M+1}(K_i - 1) = K - M - 1$), i.e., total removal of $K - 1$, then the digraph becomes acyclic. Thus

$$\mathsf{MAIS}(D_K) \geq (N' - K + 1). \qquad (26)$$

It follows from (1), (25), and (26) that $\mathsf{MAIS}(D_K) = N' - K + 1 = \ell_{\mathsf{ICC}}(D_K)$. Thus $\ell_{\mathsf{ICC}}(D_K) = \beta(D_K) = \beta_t(D_K) = N' - K + 1$. ∎

## Appendix D

### Proof of Proposition 5

*Proof:* Let $D_3$ be a 3-IC structure having an inner-vertex set $V_1 = \{a, b, c\}$. Now we prove that any non-inner vertices of $D_3$ belonging to an I-path could not contribute to form a cycle including only non-inner vertices of $D_3$. To show this, we start by picking two inner vertices $a$ and $b$, and the I-path $P_{a \to b}(D_3)$. We assume that the I-path includes $n \geq 0$ non-inner vertices, i.e., $P_{a \to b}(D_3) = \langle a, v_1, v_2, \ldots, v_n, b \rangle$. Further, let $\{v_1, v_2, \ldots, v_n\}$ be denoted by $V_n$. If $n = 0$, then $V_n = \emptyset$ for $P_{a \to b}(D_3)$. Now for $n \geq 1$, we have the following in $D_3$:

- $P_{b \to a}(D_3)$, $P_{c \to a}(D_3)$ and $P_{b \to c}(D_3)$ cannot contain any vertex in $V_n$ of $P_{a \to b}(D_3)$ because of the following:

    1) If $P_{b \to a}(D_3)$ or $P_{c \to a}(D_3)$ contains any vertex $v \in V_n$, then $P_{a \to v}(D_3)$ (part of $P_{a \to b}(D_3)$) and $P_{v \to a}(D_3)$ (part of $P_{b \to a}(D_3)$ or $P_{c \to a}(D_3)$) form an I-cycle at $a$. The existence of the I-cycle in $D_3$ contradicts the definition of an IC structure.

    2) If $P_{b \to c}(D_3)$ contains any vertex $v \in V_n$, then $P_{b \to v}(D_3)$ (part of $P_{b \to c}(D_3)$) and $P_{v \to b}(D_3)$ (part of $P_{a \to b}(D_3)$) form an I-cycle at $b$. The existence of the I-cycle in $D_3$ contradicts the definition of an IC structure.

- One can verify that only the remaining I-paths $P_{a \to c}(D_3)$ and $P_{c \to b}(D_3)$ can contain vertices in $V_n$ without forming an I-cycle in $D_3$.

Now for $P_{a \to b}(D_3)$ and $P_{a \to c}(D_3)$, these two I-paths must form a directed rooted tree $T_a$ with the root vertex $a$ (by the definition of an IC structure). Thus any set of vertices and arcs from these two I-paths alone could not form a cycle including only non-inner vertices. Furthermore, if $P_{a \to b}(D_3)$ and $P_{a \to c}(D_3)$ contain some common non-inner vertices, then let a vertex $v_i \in V_n$, which is in $T_a$, be the vertex from where $T_a$ branches to $b$ and $c$ (refer to Fig. 10a).

For $P_{a \to b}(D_3)$ and $P_{c \to b}(D_3)$, any set of vertices and arcs from these two I-paths alone could not form a cycle including only non-inner vertices because of the following:

- If they contain some common non-inner vertices, then let $v_j \in V_n$ be the first vertex where $P_{a \to b}(D_3)$ and $P_{c \to b}(D_3)$ meet each other (refer to Fig. 10b). Considering any set of vertices and arcs belonging only to these two I-paths, a cycle including only non-inner vertices can form only if a part of $P_{c \to b}(D_3)$ contributes to form a path from $v_j$ to $v_i \in V_n$ for some $i \in \{1, 2, \ldots, j-1\}$. This path is not possible because multiple I-paths would be created from $a$ or $c$ to $b$ in $D_3$ (contradiction of the definition of an IC structure). In fact, both the I-paths have the same destination, i.e., $b$, so there must be only one path from $v_j$ to $b$ common in both of them (to avoid any multiple I-paths).

For $P_{a \to c}(D_3)$ and $P_{c \to b}(D_3)$, if these I-paths contain a non-inner vertex $v$ in common, then $P_{c \to v}(D_3)$ (part of $P_{c \to b}(D_3)$) and $P_{v \to c}(D_3)$ (part of $P_{a \to c}(D_3)$) forms an I-cycle at $c$ (contradiction to the definition of an IC structure). Thus $P_{a \to c}(D_3)$ and $P_{c \to b}(D_3)$ cannot contain any vertex in common except $c$. Consequently, considering $P_{a \to b}(D_3)$, $P_{a \to c}(D_3)$ and $P_{c \to b}(D_3)$ (along with assumptions that there are some common non-inner vertices in (i) $P_{a \to b}(D_3)$ and $P_{a \to c}(D_3)$, and (ii) $P_{a \to b}(D_3)$ and $P_{c \to b}(D_3)$), we end up with the structure as shown in Fig. 10c, where $j > i$ and $i, j \in \{1, 2, \ldots, n\}$. This structure contradicts one of the necessary conditions[5] for any vertex in $V_n$ to contribute to form a cycle including only non-inner vertices in $D_3$. Thus there is no vertex in $V_n$ to contribute to form a cycle including only non-inner vertices in $D_3$.

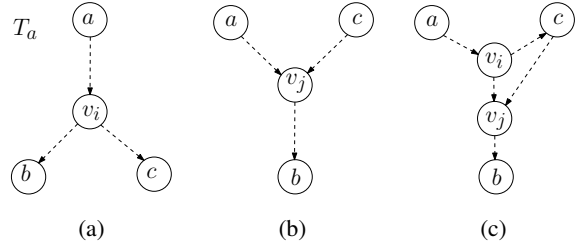

Fig. 10. (a) $P_{a \to b}(D_3)$ and $P_{a \to c}(D_3)$ forming a tree $T_a$, and the tree branches at vertex $v_i$ to go to $b$ and $c$ in $D_3$, (b) $P_{a \to b}(D_3)$ and $P_{c \to b}(D_3)$ having some vertices in $V_n$ in common, and these two paths first meet at vertex $v_j$, and (c) $P_{a \to b}(D_3)$, $P_{a \to c}(D_3)$ and $P_{c \to b}(D_3)$ together in $D_3$. The dashed lines indicate paths.

Due to symmetry, the result (any non-inner vertices of $D_3$ belonging to $P_{a \to c}(D_3)$ could not contribute to form a cycle including only non-inner vertices of $D_3$) implies similarly for non-inner vertices belonging to any of the I-paths in $D_3$. Therefore, there is no cycle among the non-inner vertices in $D_3$.

By Theorem 3, for $D_3$, we get $\ell_{\mathsf{ICC}}(D_3) = \beta(D_3) = \beta_t(D_3)$. The proof is straightforward for 2-IC structure (which is a cycle) and 1-IC structure (which is a single vertex). ∎

## Appendix E

## Proof of Theorem 4

In this section, firstly, for every minimal partial clique $D'$ with $\delta^+(D') \in \{0, 1, 2, |V(D')| - 1\}$, we prove that there exists an IC structure within it such that both of the schemes (partial-clique-cover and ICC) provide the same savings. Secondly, we conjecture that the result is valid in general (this is the main reason that results the Conjecture 2). The summary is depicted in Table III. Finally, we prove the theorem.

---

[5]There should be an incoming path to a vertex in $V_n$ (let this vertex be $v_a$) and an out-going path from a vertex in $V_n$ (let this vertex be $v_b$), for some $a \leq b$, where $a, b \in \{1, 2, \ldots, n\}$. One can easily verify this necessary condition.

We prove some lemmas (Lemmas 6, 7 and 8) that capture the properties of the minimal partial clique $D'$.

*Lemma 6:* A minimal partial clique $D'$ with $|V(D')| \geq 2$ vertices and the minimum out-degree $\delta^+(D') = 1$ is a cycle, and both the partial-clique-cover scheme and the ICC scheme achieve savings of one.

*Proof:* The properties of $D'$ with $|V(D')| \geq 2$ and $\delta^+(D') = 1$ are as follows: In $D'$, (i) as $\delta^+(D') = 1$, the partial-clique-cover scheme provides the savings of one by Proposition 6, (ii) further partitioning could not provide any savings because it is a minimal-partial-clique, and (iii) there exists at least a cycle because an acyclic digraph should have a sink vertex (a vertex with out-degree zero), and that does not exist since $\delta^+(D') = 1$. Thus $D'$, which is a minimal partial clique, having all of the above properties is a cycle. From Theorem 2, a cycle is a 2-IC structure, and the ICC scheme provides savings of one by (7). Note that such $D'$ with $|V(D')| = 2$ is also a clique of order two. ∎

Now we define some terms.

*Definition 22 (Figure-of-eight):* Two directed cycles $C_1$ and $C_2$ intersecting only at one vertex, say $u$, such that $V(C_1) \cap V(C_2) = \{u\}$, is a figure-of-eight structure at $u$.

*Lemma 7:* A minimal partial clique $D'$ with $|V(D')| \geq 1$ vertices and the minimum out-degree $\delta^+(D') = |V(D')| - 1$, is a clique of order $|V(D')|$, and $D'$ has the savings of $|V(D')| - 1$ from both the partial-clique-cover scheme and the ICC scheme.

*Proof:* In $D'$, if $\delta^+(D') = |V(D')| - 1$, then each vertex has an out-going arc to every other vertex, which is a clique by definition. Thus the partial-clique-cover scheme provides savings of $|V(D')| - 1$ by Proposition 6. From Theorem 2, a clique is a $|V(D')|$-IC structure, and the ICC scheme provides savings of $|V(D')| - 1$ by (7). ∎

*Lemma 8:* In any minimal partial clique $D'$ with $|V(D')| \geq 3$ vertices and $\delta^+(D') = 2$, there exists a 3-IC structure, and $D'$ has the savings of two from both the partial-clique-cover scheme and the ICC scheme.

We first define some terms and prove some lemmas (Lemmas 9, 10, 11 and 12) that will help to prove the Lemma 8.

*Definition 23:* A path in the forward direction indicates a path from a vertex $i$ to any vertex $j$ such that $j > i$, and a path in the reverse direction indicates a path from a vertex $j$ to any vertex $i$ such that $i < j$. For simplicity, we refer a *forward path* to a path in the forward direction, and a *reverse path* to a path in the reverse direction.

TABLE III

SMALL CAPS: Partial cliques and IC structures from the savings perspective

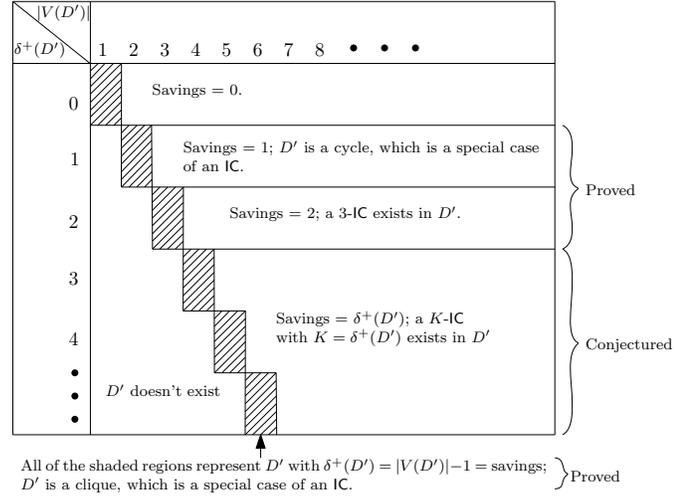

*Definition 24 (Farthest path):* Consider a sequence of vertices labeled in an increasing order such that there are multiple paths from a vertex $i$ in the sequence to other vertices in the sequence. Among those paths from $i$, the path to the vertex with the largest label is called the farthest path from vertex $i$. For example, let $1, 2, \ldots, 9$ be a sequence of vertices in an increasing order, and $1$ has paths to vertices $3, 5$ and $9$. The path from from $1$ to $9$ is the farthest path from $1$.

For the remainder of this section, we consider the following:

- A minimal partial clique $D'$ with $|V(D')| \geq 3$ number of vertices and $\delta^+(D') = 2$ has at least one cycle by Lemma 9. Without loss of generality, denote the cycle by $C$, and its vertices by $V(C) = \{1, 2, \ldots, p\}$. For simplicity, vertices of $C$ are labeled in an increasing order as shown in Fig. 12a.

- For $a, b \in V(C)$, $P_{a \to b}(C)$ denotes a path from $a$ to $b$ including only vertices and arcs of $C$, and $P'_{a \to b}(D')$ denotes a path from $a$ to $b$ consisting of arcs and vertices outside $C$ except vertices $a$ and $b$.

- For the vertices in $V(C)$ (see Fig. 12a), vertex $1$ (the first vertex of $C$) has only forward paths to other vertices of $C$ (Definition 23), and vertex $p$ (the last vertex of $C$) has only reverse paths to other vertices of $C$ (Definition 23). Let $i \in V(C) \setminus \{1, p\}$ be the first vertex in the vertex sequence that has a reverse path to any vertex in $V(C)$, i.e., there exist a $P'_{i \to j}(D')$ such that $j \in \{1, 2 \ldots, i-1\}$ and all vertices in $\{1, 2 \ldots, i-1\}$ have no reverse path.

*Lemma 9:* In a minimal partial clique $D'$ with the minimum out-degree $\delta^+(D') = 2$, and $|V(D')| \geq 3$ vertices;

1) there exists at least one cycle, $C$, and

2) for a vertex $v \in V(C)$, one of its out-going arcs (beside the arc in $C$) contribute to form a path that always returns to some vertex in $C$.

*Proof:* In $D'$ with $|V(D')| \geq 3$ vertices and $\delta^+(D') = 2$, there exists at least one cycle, denoted $C$. This is because if any $D'$ is acyclic, then it must contain at least one sink vertex, and there cannot be sink vertices in $D'$ since $\delta^+(D') = 2$.

Assume that vertices $u, v \in V(C)$ such that there is an arc from $u$ to $v$ in $C$. Now in $D'$, $u$ has out-degree $d^+_{D'}(u) \geq 2$, and the next out-going arc of $u$ (other than the arc in $C$)

(i) goes to any vertex in $V(C) \setminus \{v\}$, or

(ii) contributes a path, say $Q$, which includes at least one vertex outside $C$.

Consider the terminal vertex of the path $Q$ to be $z$ such that $z \in V(D') \setminus V(C)$. Vertex $z$ cannot be a sink vertex because $d^+_{D'}(z) \geq 2$, and it must contribute to form a path. Since there are no disjoint cycles (all cycles are connected, otherwise any two disjoint cycles in $D'$ provides savings of two and such $D'$ is not a minimal partial clique with $\delta^+(D') = 2$), $z$ must have a return path to a vertex in $V(C)$. ∎

Analyzing $D'$ by considering the out-going paths from vertices of $C$, we get the following lemmas.

*Lemma 10:* Any path $P'_{k \to m}(D')$ for $k \in \{j+1, j+2, \ldots, i-1\}$, $m \in V(C) \setminus \{i\}$ and a path $P'_{i \to j}(D')$ are disjoint.

*Proof:* We had considered that vertex $i$ of $C$ is the first vertex having a reverse path. Now if any path $P'_{k \to m}(D')$ meets path $P'_{i \to j}(D')$ at some vertices, then there exists a reverse path from vertex $k$ to $j$. This is not possible because such $k$ would have been the first vertex in the vertex sequence that has a reverse path (contradiction). ∎

*Lemma 11:* Any farthest path from $j$, $P'_{j \to k}(D')$, for $k \in \{j+1, j+2, \ldots, p\}$ and a path $P'_{m \to q}(D')$ for $m \in \{j+1, j+2, \ldots, k-1\}$, $q \in \{k+1, k+2, \ldots, p\}$ are disjoint.

*Proof:* If the farthest path from $j$, $P'_{j \to k}(D')$, meets any path $P'_{m \to q}(D')$ at some vertices, then there exists $P'_{j \to q}(D')$. This is not possible; otherwise, path $P'_{j \to q}(D')$ would have been the farthest path (contradiction). ∎

*Lemma 12:* If any path $P'_{j \to k}(D')$ and a path $P'_{m \to j}(D')$ for any $m \neq k$ meet at some common vertices except $j$, then there exists a figure-of-eight at $j$.

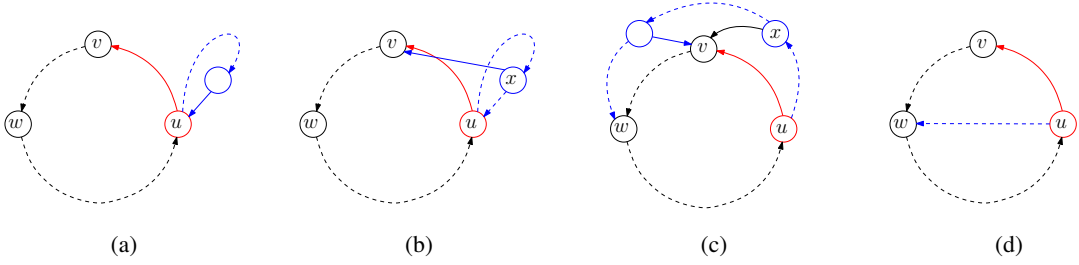

Fig. 11. The possible cases of out-going arcs for vertex $u \in V(C)$. The paths are indicated by dashed lines, and the arcs are indicated by solid lines.

*Proof:* If $j = k$ or $m = j$, then we get a figure-of-eight structure at $j$. Two closed paths at $j$ will be (i) $P'_{j \to j}(D')$, and (ii) $C$. If $j \neq k \neq m$, then let $x$ be the common vertex which is nearest to $j$ in these two paths $P'_{j \to k}(D')$ and $P'_{m \to j}(D')$. Now we get a figure-of-eight structure at $j$. Two closed paths at $j$ will be (i) $P_{j \to x}(D')$ (part of $P'_{j \to k}(D')$), $P_{x \to j}(D')$ (part of $P'_{m \to j}(D')$), and (ii) $C$. These closed paths are disjoint except $j$. ∎

Now we prove Lemma 8.

*Proof of Lemma 8:* The proof is done by the detailed structural analysis of the minimal partial clique $D'$. We divide the proof into two parts: In Part-I, we prove that $D'$ has a figure-of-eight structure (see Definition 22) at a vertex, and in Part-II, we prove there exist a 3-IC structure within $D'$ having a figure-of-eight structure.

(Part-I) Consider the cycle $C$ in $D'$ has $p \geq 2$ vertices. Now based on $C$, there are two cases in $D'$, and those are (i) $C$ with $p = 2$, i.e., $C$ is a clique of size two, and (ii) $C$ with $p \geq 3$. We assume $u$ and $v$ are the two distinct vertices belonging to $V(C)$ in such a way that there exist a directed arc from $u$ to $v$. In $D'$, $d^+_{D'}(u) \geq 2$ and the next out-going arc of $u$ goes out of $C$, and contributes to form a path, say $Q$ that returns to some vertex in $V(C)$ by Lemma 9.

(Case (i)) The cycle $C$ is a clique of size two. Therefore, the path $Q$ returns to $C$ either at $u$ or $v$. Now we get a figure-of-eight structure at $u$ if the path returns to $u$, otherwise it returns to $v$. For the latter case, we get a new cycle that includes the path $Q$ (which includes $u$, some vertices other than $v$), vertex $v$ and the arc from $v$ to $u$ (i.e., arc of the cycle $C$). The new cycle has more than two vertices, thus this ends up with case (ii).

(Case (ii)) On the basis where path $Q$ returns, we have the following sub-cases: $Q$ returns to (ii-A) $u$ (see in Fig. 11a) providing a figure-of-eight structure, (ii-B) $v$ (see in Fig. 11b and 11c), and (ii-C) some vertex $w \in V(C) \setminus \{u, v\}$ (see in Fig. 11d).

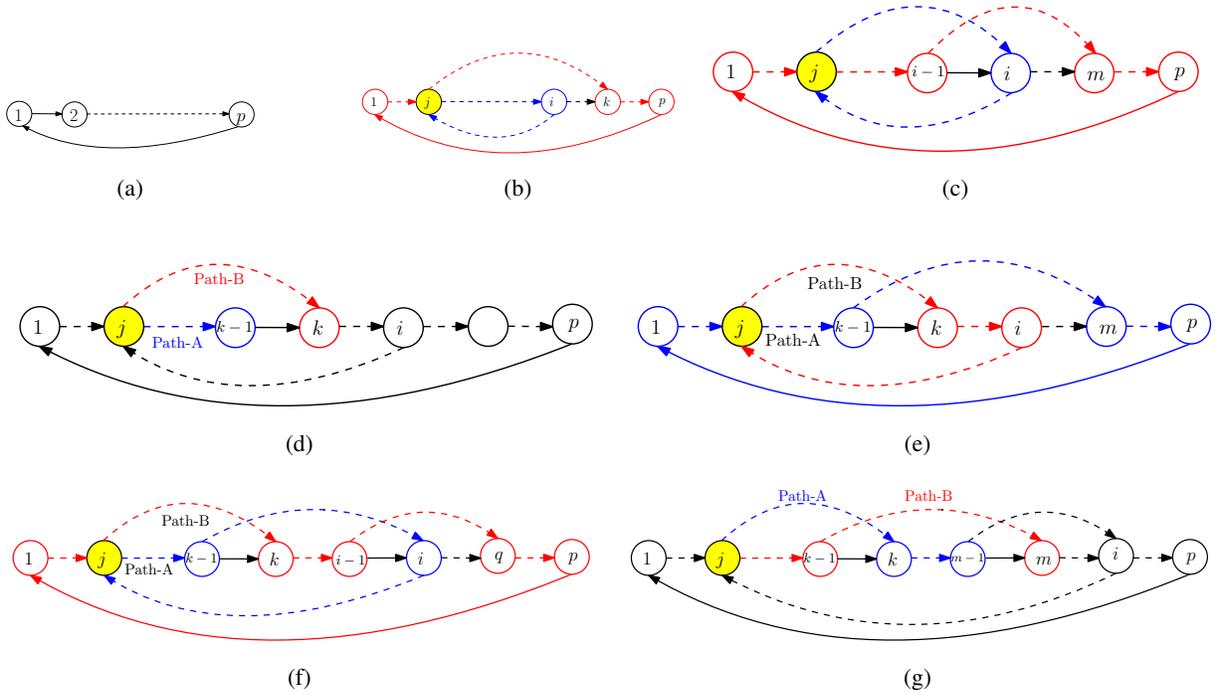

Fig. 12. (a) Cycle $C$ and its vertices, (b) figure-of-eight at vertex $j$ with $P'_{j \to k}(D')$ for $k \in \{i+1, i+2, \ldots, p\}$, (c) figure-of-eight at vertex $j$ with $P'_{j \to i}(D')$, (d) illustration of path-A and path-B, (e) figure-of-eight at vertex $j$ with path-A, path-B and $P'_{k-1 \to m}(D')$ for $m \in \{i+1, i+2, \ldots, p\}$, (f) figure-of-eight at vertex $j$ with path-A, path-B and $P'_{k-1 \to i}(D')$, and (g) illustration of updating path-A and path-B. The paths are indicated by dashed lines, and the arcs are indicated by solid lines.

For sub-case (ii-B), we have a vertex $x \in V(Q) \setminus V(C)$ such that there is a direct arc from $x$ to $v$. Since there are no disjoint cycles, the next out-going arc of $x$ (besides the arc from $x$ to $v$) contributes to form a path to a vertex in $V(C)$. If $x$ has path to $u$, then we get a figure-of-eight at $u$ (shown in Fig. 11b), otherwise we have the following:

a) $x$ has a path to some $w \in V(C) \setminus \{u, v\}$, so this ends up with sub-case (ii-C) (shown in Fig. 11d), or

b) $x$ has a path to $v$, but for this case, we have another path (shown in Fig. 11c) from $u$ to $v$ (beside the direct arc from $u$ to $v$, and path including the direct arc from $x$ to $v$), and in this path, we can repeat sub-case (ii-B) by considering the predecessor of $v$ in place of $x$. Since the number of vertices in $D'$ is finite, sub-case (ii-B) either ends up with a figure-of-eight structure or sub-case (ii-C).

For sub-case (ii-C), note that we have a path $Q$, which is disjoint from $C$ except the first and the last vertices, which starts from any vertex $u \in V(C)$ and returns to some vertex $w \in V(C) \setminus \{u, v\}$.

We start analyzing the sub-case (ii-C) in $D'$ considering vertices in $V(C)$. For the vertices in $V(C)$ (see Fig. 12a), in a sequential order starting from vertex 1, we track their out-going paths (which may include vertices not in $V(C)$) to other vertices in $V(C)$. For this sub-case, we know that there exist a $P'_{i \to j}(D')$ such that $j \in \{1, 2 \ldots, i-1\}$ and all vertices in $\{1, 2 \ldots, i-1\}$ have only forward paths. Now we consider out-going paths from $j$ (the paths are always forward paths), and get the following subsub-cases:

1) If the farthest path from $j$ is $P'_{j \to k}(D')$ for some $k \in \{i+1, i+2, \ldots, p\}$, then there exists a figure-of-eight at $j$. Two closed paths at $j$ will be (i) $P_{j \to i}(C)$, $P'_{i \to j}(D')$, and (ii) $P'_{j \to k}(D')$, $P_{k \to j}(C)$ (refer Fig. 12b). If $P'_{j \to k}(D')$ and $P'_{i \to j}(D')$ are disjoint except $j$, then by recalling the definition of $P_{a \to b}(C)$ and $P'_{a \to b}(D')$ for any $a, b \in V(C)$, one can show that the two closed paths are disjoint except $j$. Otherwise, $P'_{j \to k}(D')$ and $P'_{i \to j}(D')$ are not disjoint except $j$, and by Lemma 12, one can find a figure-of-eight at $j$.

2) If the farthest path from $j$ is $P'_{j \to k}(D')$ for $k = i$, then there exists a figure-of-eight at $j$. Two closed paths at $j$ will be (i) $P'_{j \to i}(D')$, $P'_{i \to j}(D')$, and (ii) $P_{j \to i-1}(C)$, $P'_{i-1 \to m}(D')$ for $m \in \{i+1, i+2, \ldots, p\}$ (one of the forward paths of $i-1$), $P_{m \to j}(C)$ (refer Fig. 12c). Using Lemmas 10 and 11, and recalling the definition of $P_{a \to b}(C)$ and $P'_{a \to b}(D')$ for any $a, b \in V(C)$, one can show that these two closed paths are disjoint except $j$.

3) Otherwise, the farthest path from $j$ is $P'_{j \to k}(D')$ for some $k \in \{j+2, j+3, \ldots, i-1\}$. Starting from $j$, we have at least two forward paths to some vertices in $V(C)$ such that these paths are disjoint except $j$. We assume these paths are path-A (the path from $j$ to $k-1$) and path-B (the path from $j$ to $k$). Paths $P_{j \to k-1}(C)$ and $P'_{j \to k}(D')$ will be path-A and path-B respectively (refer Fig. 12d). Now considering the out-going paths from $k-1$ into account (the paths are always forward paths), we get the following subsubsub-cases:

   (a) If the farthest path from $k-1$ is $P'_{k-1 \to m}(D')$ for some $m \in \{i+1, i+2, \ldots, p\}$, then there exists a figure-of-eight at $j$. Two closed paths at $j$ will be (i) path-A, $P'_{k-1 \to m}(D')$, $P_{m \to j}(C)$, and (ii) path-B, $P_{k \to i}(C)$, $P'_{i \to j}(D')$ (refer Fig. 12e). Using Lemmas 10 and 11, and recalling the definition of $P_{a \to b}(C)$ and $P'_{a \to b}(D')$ for any $a, b \in V(C)$, one can show that these two closed paths are disjoint except $j$.

   (b) If the farthest path from $k-1$ is $P'_{k-1 \to m}(D')$ for $m = i$, then there exists a figure-of-eight at $j$. Two closed paths at $j$ are (i) path-A, $P'_{k-1 \to i}(D')$, $P'_{i \to j}(D')$, and (ii) path-B, $P_{k \to i-1}(C)$, $P'_{i-1 \to q}(D')$ for $q \in \{i+1, i+2, \ldots, p\}$ (one of the forward paths of $i-1$), $P_{q \to j}(C)$ (refer Fig. 12f). If $P'_{j \to k}(D')$ and $P'_{i \to j}(D')$ are disjoint except $j$, then by using Lemmas 10 and 11, and recalling the definition of $P_{a \to b}(C)$ and $P'_{a \to b}(D')$ for any $a, b \in V(C)$, one can show that the two closed paths are disjoint except $j$. Otherwise, $P'_{j \to k}(D')$ and $P'_{i \to j}(D')$ are not disjoint except $j$, and by Lemma 12, one can find a figure-of-eight at $j$.

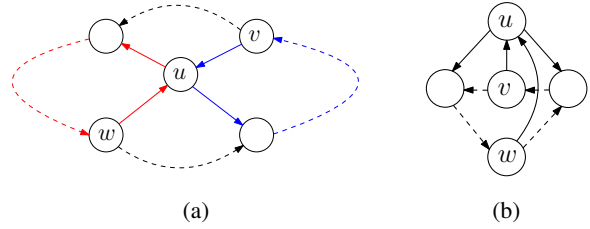

(a)                                    (b)

Fig. 13. (a) Structure of $D'$ with the minimum out-degree two, and (b) rearranging $D'$ to get an IC structure. The paths are indicated by dashed lines, and the arcs are indicated by solid lines.

(c) Otherwise, the farthest path from $k-1$ is $P'_{k-1 \to m}(D')$ for some $m \in \{k+1, k+2, \ldots, i-1\}$. The union of path-B and $P_{k \to m-1}(C)$ give new path-A and we update $k-1 = m-1$. Similarly, the union of path-A and $P'_{k-1 \to m}(D')$ give new path-B and we update $k = m$ (Refer Fig. 12g). Considering new path-A, new path-B, and updating $k = m$, we repeat the subsubsub-cases of subsub-case 3. During this iteration, if we get subsubsub-case (a) or (b), then there is a figure-of-eight. Otherwise, we have subsubsub-case (c), where $k$ strictly increase to a value up to $i-1$. However, when $k$ reaches $i-1$, we must have either subsubsub-case (a) or (b).

Thus for all cases, $D'$ with $|V(D')| \geq 3$ and $\delta^+(D') = 2$ has a figure-of-eight structure at a vertex.

(Part-II) Without loss of generality, we consider a figure-of-eight structure at vertex $u$ (two intersecting cycles at $u$ are indicated by red and blue colors in Fig. 13a).

Now for the figure-of-eight structure at $u$ in $D'$ (see Fig. 13a), consider a vertex $v$ in a cycle $C_1$ (indicated in blue color), and a vertex $w$ in another cycle $C_2$ (indicated in red color) in such a way that both $v$ and $w$ have direct arcs going to $u$. One of the out-going arcs of $v$ must contribute to form a path, say $Q$, returning to a vertex in $V(C_2) \setminus \{u\}$. This is because all other cases are not possible;

- $Q$ cannot return to any vertex in $V(C_1) \setminus \{u, v\}$, otherwise a disjoint cycle to $C_2$ will be created.

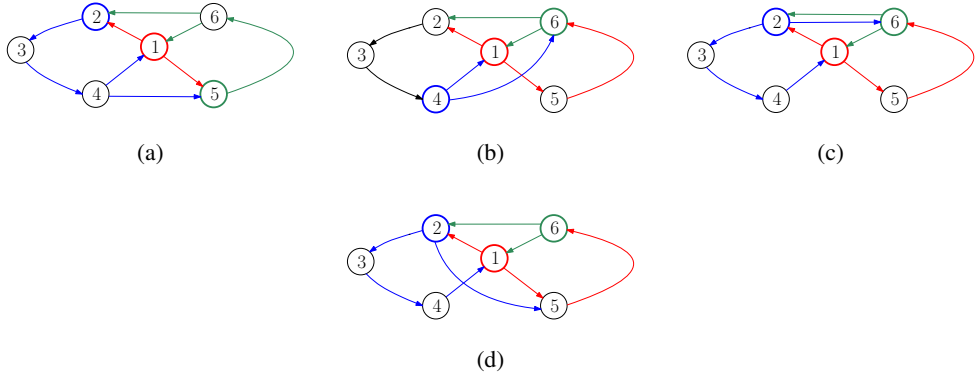

Fig. 14. Vertices in red, blue and green colors are three internal vertices for the 3-IC in each of the cases.

- $Q$ can return to $u$, but for this case, we have another path from $v$ to $u$ (beside path including direct arc from $v$ to $u$), and in this path, we can repeat the case by considering the predecessor of $u$ in place of $v$. Thus this case ends up with the same consideration as of the cycle $C_1$ with the vertex $v$ having direct arc to the vertex $u$.

- $Q$ must return to some vertex in $V(C_1)$ and $V(C_2)$ due to Lemma 9.

Now in a similarly way, one of the out-going arcs of $w$ must also contribute a path to a vertex in $V(C_1)$ other than $u$. Rearrange $D'$ to get the structure in Fig. 13b. Now consider $V_I = \{u, v, w\}$. We can see that any vertex in $V_I$ has only one I-path each to other two vertices in $V_I$ with no I-cycles. Thus a 3-IC exists in $D'$ (for example, see Fig. 14). Consecutively, the ICC scheme provides savings of two by (7). Again, for $D'$, the partial-clique-cover scheme provides savings of two by Proposition 6. ∎

For all minimal partial cliques that we have analyzed, there exists an $(\delta^+(D') + 1)$-IC structure within each minimal partial clique having the minimum out-degree $\delta^+(D')$. We conjecture that this holds in general (the following conjecture is not a part of the proof of the Theorem 4, but this provides an intuition about the Conjecture 2).

*Conjecture 3:* For any minimal partial clique $D'$, both the partial-clique-cover scheme and the ICC scheme provides the same savings.

Now we prove Theorem 4.

*Proof of Theorem 4:* Given a digraph $D$, if minimal partial cliques, partitioning the digraph to provide partial-clique number, have $\delta^+(D') \in \{0, 1, 2, |V(D)| - 1\}$, then using the ICC scheme on each of the minimal partial cliques, we achieve $\ell_{PC}(D)$ by using Lemmas 6, 7 and 8, and

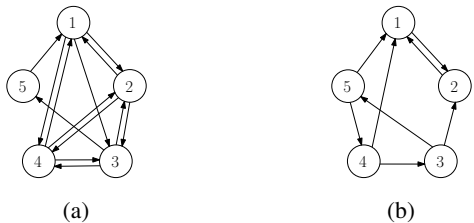

Fig. 15. (a) The digraph, denoted $D_1$, has $\ell_{\mathsf{PC}}(D_1) = 3$ (considering possible sub-digraphs with vertex sets; $\{2, 3, 4\}$ forming a clique, $\{1\}$ and $\{5\}$ forming two cliques each with single vertex), but $\ell_{\mathsf{ICC}}(D_1) = 2$ (considering sub-digraphs with vertex sets; $\{1, 2, 3, 4, 5\}$ forming an IC structure with 1, 2, 3 and 4 as inner vertices), and (b) the digraph, denoted $D_2$, has $\ell_{\mathsf{PC}}(D_2) = 3$ (considering possible sub-digraphs with vertex sets; $\{1, 2\}$ forming a clique, and $\{3, 4, 5\}$ forming a cycle), and $\ell_{\mathsf{ICC}}(D_2) = 3$ (considering sub-digraphs with vertex sets; $\{1, 2\}$ forming an IC structure, and $\{3, 4, 5\}$ forming another IC structure).

considering zero savings for $\delta^+(D') = 0$ from any schemes. The ICC scheme may produce a shorter index code by considering a different partitioning of the digraph (for example, see Fig. 15a and 15b). ∎

*Example 5:* This example illustrates that for the class of digraphs stated in Theorem 4, the ICC scheme performs as least as well as the partial-clique-cover scheme. Consider two digraphs that are depicted in Fig. 15. The digraph in Fig. 15a has more savings from the ICC scheme than that obtained from the partial-clique-cover scheme, and the digraph in Fig. 15b has equal savings from both schemes.

## APPENDIX F

Note that $K$ is an even integer greater than 2, $N = \frac{3K}{2}$, and $i, j \in \{1, 2, \ldots, \frac{K}{2}\}$ such that $j \neq i$.

*Proposition 8:* For any digraph $D$ of the Class A, the index codelength obtained from the partial-clique-cover scheme and the ICC scheme are $K$ and $\frac{K}{2} + 1$ respectively, i.e., $\ell_{\mathsf{PC}}(D) = K$ and $\ell_{\mathsf{ICC}}(D) = \frac{K}{2} + 1$.

Firstly, we prove some lemmas.

*Lemma 13:* Any minimal partial clique $D'$ in a digraph of the Class A provides savings less than or equal to two, i.e., $S(D') \leq 2$.

*Proof:* If any minimal partial clique $D'$ in $D$ of the Class A includes a vertex $K + i$ for any $i \in \{1, 2, \ldots, \frac{K}{2}\}$, then $\delta^+(D') \leq 2$. This is because the out-degree of any vertex $K + i$ in $D$ is two, i.e., $d_D^+(K + i) = 2$ by construction. Now any minimal partial clique without any vertex in $\{K + 1, K + 2, \ldots, \frac{3K}{2}\}$ has $\delta^+(D') \leq 1$. This is because for any vertex in $\{1, 2, \ldots, K\}$, only one out-neighbor is in this set, and the rest are in $\{K+1, K+2, \ldots, \frac{3K}{2}\}$. So, any minimal partial clique in the digraph can have $\delta^+(D') \leq 2$. Now by proposition 6, we know that $S(D') = \delta^+(D')$. ∎

*Lemma 14:* A digraph $D$ of the Class A has no minimal partial clique with $\delta^+(D') = 2$.

*Proof:* In $D$, we try to construct a minimal partial clique $D'$ that has $\delta^+(D') = 2$ (we do not need to consider $D'$ having $\delta^+(D') > 2$ because of Lemma 13) by starting from $D'$ with only one vertex and then adding vertices into its vertex set in such a way that we can obtain $\delta^+(D') = 2$.

We start from any vertex $K+i$ for an $i \in \{1, 2, \ldots, \frac{K}{2}\}$ (we will show a similar result if we start from some vertex $i$). Let $V(D') = \{K+i\}$ be the vertex set of $D'$. Now we include both of the two out-neighbor vertices of $K+i$, i.e., vertices in $N_D^+(K+i)$ in $V(D')$. If we include only one out-neighbor vertex of $K+i$ in $V(D')$, then $d_{D'}^+(K+i) = 1$ resulting $\delta^+(D') = 1$. The new vertex-induced sub-digraph $D'$ has $V(D') = \{K+i, 2i-1, 2i\}$, and $\delta^+(D') = 1$ because $d_{D'}^+(2i-1) = d_{D'}^+(2i) = 1$. To get a minimum out-degree of two, we must include another vertex in $V(D')$ from $\{K+1, K+2, \ldots, \frac{3K}{2}\} \setminus \{i\}$. By symmetry, it does not make any difference which vertex to add. We arbitrarily choose $K+j$, for some $j \in \{1, 2, \ldots, \frac{K}{2}\} \setminus \{i\}$. Now the new vertex-induced sub-digraph $D'$ has $V(D') = \{K+i, 2i-1, 2i, K+j\}$ for a $j$. Here $\delta^+(D') = 0$ because $d_{D'}^+(K+j) = 0$. We further include all vertices in $N_D^+(K+j)$ in $V(D')$ because $d_D^+(K+j) = 2$, and if we include only one of its out-neighbor vertices, then $\delta^+(D') = 1$. The new $D'$ has $V(D') = \{K+i, 2i-1, 2i, K+j, 2j-1, 2j\}$ and $\delta^+(D') = 2$. Further, including any vertex set in $V(D')$ could not increase $\delta^+(D')$ because any $D'$ must have $\delta^+(D') = S(D') \leq 2$ (Lemma 13). If we start building $D'$ with some $m \in \{1, 2, \ldots, K\}$, we will end up with a sub-digraph (i) that includes $\{m, K+i, 2i-1, 2i, K+j, 2j-1, 2j\}$ for some $i, j \in \{1, 2, \ldots, \frac{K}{2}\}$ where $i \neq j \neq m$, or (ii) that includes $\{2i-1, 2i, K+j, 2j-1, 2j, K+i\}$ for some $i, j \in \{1, 2, \ldots, \frac{K}{2}\}$ where $i \neq j$ and $m = 2i$ or $2i-1$. Altogether, any minimal partial clique with $\delta^+(D') = 2$ must contain $\{K+i, 2i-1, 2i, K+j, 2j-1, 2j\}$ for some $i, j \in \{1, 2, \ldots, \frac{K}{2}\}$ where $i \neq j$. However, $D'$ is not the minimal because by simply considering two partial cliques among the vertices in $D'$, i.e., partial cliques with vertex sets $\{2i-1, 2i\}$ and $\{2j-1, 2j\}$, we get savings of two (one in each). ∎

*Proof of Proposition 8:* For a minimal partial clique $D'$ in $D$ of the Class A, Lemma 13 provides $\delta^+(D') \leq 2$, and Lemma 14 proves that there exists no minimal partial clique with $\delta^+(D') = 2$. Thus the minimal partial cliques in a digraph $D$ of the Class A are only cycles. For this case, the cycle-cover scheme and the partial-clique-cover scheme for $D$ are the same. Since any cycle must include two vertices from $\{1, 2, \ldots, K\}$, $\ell_{\mathsf{PC}}(D) \geq \frac{3K}{2} - \frac{K}{2} = K$. The minimal partial cliques with vertex sets $\{1, 2\}, \{3, 4\}, \ldots, \{K-1, K\}, \{K+1\}, \{K+2\}, \ldots, \{N\}$ provide $\ell_{\mathsf{PC}}(D) = \frac{K}{2} + \frac{K}{2} = K$.

Consider a vertex set $V_{\mathrm{I}} = \{1, 2, \ldots, K\}$ with $K$ number of vertices. In the digraph $D$, any vertex $v \in V_{\mathrm{I}}$ has only one path each to all other vertices in $V_{\mathrm{I}} \setminus \{v\}$ such that only the first and the last vertices of each of these paths belong to $V_{\mathrm{I}}$ (i.e., I-path). Moreover, there is no I-cycle at any vertex in $V_{\mathrm{I}}$. Thus this forms an IC structure with inner-vertex set $V_{\mathrm{I}}$. Now from (7), $\ell_{\mathsf{ICC}}(D) = N - K + 1$. Thus $\ell_{\mathsf{ICC}}(D) = 3 \times (\frac{K}{2}) - K + 1 = \frac{K}{2} + 1$. ∎

# Corrections to "Interlinked Cycles for Index Coding: Generalizing Cycles and Cliques"

Chandra Thapa, Lawrence Ong, and Sarah J. Johnson

We provide a correction to [1] in response to an error reported by M. B. Vaddi and B. S. Rajan [2]. To this effect, we add one extra condition for the definition of an IC structure on page 3696[1].

This correction is required because referring to the proof of the Lemma 3 on page 3702[2], the following statements may not be correct in general:

- Statement 1: In the digraph $D_K$, we can also obtain a directed path from $j$ which passes through $v$ (via $T_j$), and ends at the leaf vertex $c$ (via $T_i$).

- Statement 2: Thus a path $\langle j, \ldots, v, b, \ldots, d \rangle$ exists in $D_K$.

In particular, Statement 1 is valid if and only if $P_{j \to v}(T_j) \cap P_{v \to c}(T_i) = \{v\}$, i.e., the paths $P_{j \to v}(T_j)$ and $P_{v \to c}(T_i)$ have no vertex in common except $v$, where $P_{j \to v}(T_j)$ is the path from $j$ to $v$ in $T_j$, and $P_{v \to c}(T_i)$ is the path from $v$ to $c$ in $T_i$. We observe a similar problem in Statement 2.

To address this issue, we introduce an additional constraint[3] to the definition of an IC structure on page 3696[1] (stated as Condition 3 in the following). The conditions for an IC structure are modified as follows:

- *Condition 1:* There is no I-cycle in $D_K$.

- *Condition 2:* For all ordered pairs of inner vertices $(i, j)$, $i \neq j$, there is only one I-path from $i$ to $j$ in $D_K$ (this condition implies that in $D_K$, for any two inner vertices, there is exactly one cycle containing them and no other inner vertices).

- *[New:] Condition 3:* No cycle containing only non-inner vertices.

---

[1] page 10 of the arXiv version

[2] page 25 of the arXiv version

[3] This constraint was introduced in "Graphical Approaches to Single-Sender and Two-Sender Index Coding", PhD thesis, submitted on 24 Jan. 2018 by C. Thapa for examination.

With the introduction of Condition 3, Statements 1 and 2 are now correct, otherwise a cycle containing only non-inner vertices can be formed, which contradicts the new definition of an IC structure. For example, for Statement 1, if there exists at least another non-inner vertex, say $a$, such that $a \neq v$ and $a \in P_{j \to v}(T_j) \cap P_{v \to c}(T_i)$, then there exist $P_{a \to v}(T_j)$ and $P_{v \to a}(T_i)$, which gives a cycle including only non-inner vertices $\langle a, \dots, v, \dots, a \rangle$.

The aforementioned correction was also proposed independent and in parallel by K. V. Bharadwaj and B. S. Rajan [3].

The new definition of an IC structure always satisfies Case 1 in Theorem 3 on page 3698[4]; Case 2 (and the corresponding proof on page 3705[5]) is no longer required. So, Theorem 3 can be strengthened as follows:

*[New:] Theorem 3:* For messages of any $t \geq 1$ bits, for any $K$-IC structure $D_K$, the scalar linear index code given by the ICC scheme is optimal for $D_K$, i.e., $\ell_{\mathsf{ICC}}(D_K) = \beta(D_K) = \beta_t(D_K)$.

Now, Proposition 5 on page 3698[6] is a special case of the modified Theorem 3 and should be deleted. Also, Conjecture 1 on page 3698[7] is the same as the modified Theorem 3, so it is no longer required and should be deleted.

In Algorithm 2 on page 3705[8], the phrase

"such that $V(P_{i \to j}(D'_s)) \cap V' = \{i, j\}$ then"

is replaced by

[New:] "such that $V(P_{i \to j}(D'_s)) \cap V' = \{i, j\}$ and no cycle only among $V(P_{i \to j}(D'_s)) \setminus \{i, j\}$ then"

to take Condition 3 into account.

Besides the foregoing changes and updates, the remainder reported results and proofs are intact.

## ACKNOWLEDGMENT

The authors would like to thank Dr. Min Li for useful discussions.

---

[4]page 15 of the arXiv version

[5]page 33 of the arXiv version

[6]page 15 of the arXiv version

[7]page 16 of the arXiv version

[8]page 32 of the arXiv version